\pdfoutput=1
\documentclass[12pt,a4paper]{article}
\usepackage{jheppub}

\RequirePackage[utf8]{inputenc}

\usepackage{diagbox}

\usepackage[graphicx]{realboxes}
\usepackage{lscape}

\usepackage{amsfonts}
\usepackage{amssymb}
\usepackage{comment}
\usepackage{graphicx}
\usepackage{amsmath}
\usepackage{tabu}
\usepackage{dsfont}
\usepackage{float}
\usepackage{amsthm}
\usepackage{slashed}
\usepackage{setspace}
\usepackage[labelformat=simple]{subcaption}

\usepackage{listings}

\usepackage[utf8]{inputenc}
\usepackage{amsmath, amsthm, amssymb, amsfonts}
\usepackage[font=small, labelfont=bf]{caption}
\usepackage{amsmath, amsthm, amssymb, amsfonts}
\usepackage{multirow}
\usepackage{graphicx}
\usepackage{float}
\usepackage[numbers,sort&compress]{natbib}
\usepackage{jheppub}
\usepackage{enumerate}

\usepackage{wrapfig}
\usepackage{booktabs}

\newcommand{\nn}{\nonumber}

\newcommand{\bZ}{\mathbb{Z}}
\newcommand{\bR}{\mathbb{R}}
\newcommand{\bN}{\mathbb{N}}
\newcommand{\bP}{\mathbb{P}}
\newcommand{\cF}{\mathcal{F}}

\newcommand{\raw}{\rightarrow}
\newcommand{\cN}{\mathcal{N}}

\newcommand{\cV}{\mathcal{V}}

\newcommand{\cO}{\mathcal{O}}

\newcommand{\kom}{\, ,\quad }

\newcommand{\nc}{\newcommand}
\nc{\beq}{\begin{equation}}
\nc{\eeq}{\end{equation}}
\nc{\bea}{\begin{eqnarray}}
\nc{\eea}{\end{eqnarray}}

\allowdisplaybreaks[2]
\numberwithin{equation}{section}

\usepackage[framemethod=TikZ]{mdframed}
\mdfsetup{skipabove=\topskip, skipbelow=\topskip}

\newcounter{equ}[section]

\newcounter{Boxequ}[section]

\title{A Database of Calabi-Yau Orientifolds \\ and the Size of D3-Tadpoles}

\author[a]{Chiara Crin{\`o},}
\author[b]{Fernando Quevedo,}
\author[b]{Andreas Schachner,}
\author[a]{Roberto Valandro}

\affiliation[b]{\footnotesize DAMTP, Centre for Mathematical Sciences, Wilberforce Road, Cambridge, CB3 0WA, UK.}
\affiliation[a]{\footnotesize Dipartimento di Fisica, Universitá di Trieste, Strada Costiera 11, I-34151 Trieste, Italy and INFN, Sezione di Trieste, via Valerio 2, I-34127 Trieste, Italy}

\emailAdd{Chiara.Crino@ts.infn.it}
\emailAdd{fq201@damtp.cam.ac.uk}
\emailAdd{as2673@maths.cam.ac.uk}
\emailAdd{roberto.valandro@ts.infn.it}

\abstract{
The classification of 4D reflexive polytopes by Kreuzer and Skarke allows for a systematic construction of Calabi-Yau hypersurfaces as fine, regular, star triangulations (FRSTs).
Until now, the vastness of this geometric landscape remains largely unexplored.
In this paper, we construct Calabi-Yau orientifolds from holomorphic reflection involutions of such hypersurfaces with Hodge numbers $h^{1,1}\leq 12$.
In particular, we compute orientifold configurations for all favourable FRSTs for $h^{1,1}\leq 7$, while randomly sampling triangulations for each pair of Hodge numbers up to $h^{1,1}=12$.
We find explicit string compactifications on these orientifolded Calabi-Yaus for which the D3-charge contribution coming from O$p$-planes grows linearly with the number of complex structure and Kähler moduli.
We further  consider non-local D7-tadpole cancellation through Whitney branes.
We argue that this leads to a significant enhancement of the total D3-tadpole as compared to conventional $\mathrm{SO}(8)$ stacks with $(4+4)$ D7-branes on top of O7-planes.
In particular, before turning-on worldvolume fluxes,
we find that the largest D3-tadpole in this class occurs for Calabi-Yau threefolds with $(h^{1,1}_{+},h^{1,2}_{-})=(11,491)$ with D3-brane charges  $|Q_{\text{D3}}|=504$ for the local D7 case and  $|Q_{\text{D3}}|=6,664$ for the non-local Whitney branes case, which appears to be large enough to cancel tadpoles and allow fluxes to stabilise all complex structure moduli.
Our data is publicly available under the following link \url{https://github.com/AndreasSchachner/CY_Orientifold_database}.

}


\usepackage{todonotes}

\begin{document}

\maketitle

\bigskip

\newpage

\section{Introduction}

Within the general context of  flux compactifications in string theory, the goal of this paper is twofold: First, to discuss the size of D3-tadpoles in the presence of local and non-local D7 configurations. 
Secondly,
to generate a database of Calabi-Yau (CY)  orientifolds from reflection involutions that allows us to explicitly determine the size of D3-tadpoles in concrete models and that may have further applications.

Regarding the second goal, we provide a complete scan of type IIB orientifold models with O3/O7-planes for $h^{1,1}\leq 7$ for CY hypersurfaces obtained from the Kreuzer-Skarke (KS)  database \cite{Kreuzer:2000xy} via reflection involutions $z\rightarrow -z$ of toric coordinates.
For $8\leq h^{1,1}\leq 12$, we compute one triangulation per polytope to search for further appropriate  models.
We stress that our methods are easily applied to any triangulation of any polytope in the KS database.

The KS database \cite{Kreuzer:2000xy} has received substantial attention in recent years, especially with the advent of software developments such as \texttt{CYTools} \cite{Demirtas:2020dbm}, making geometries at $h^{1,1}>10$ readily accessible.
Our investigation complements the analysis of \cite{Altman:2021pyc,Gao:2021xbs} for exchange involutions in the KS database for $h^{1,1}\leq 6$ as well as of \cite{Carta:2020ohw} for Complete Intersection Calabi-Yaus (CICYs).
Our database contains 71,941,643 orientifolds and extends previous orientifold databases of 2,004,513 CICY orientifolds \cite{Carta:2020ohw} and 28,463 divisor exchange involutions \cite{Altman:2021pyc}.\footnote{To be more concrete,
we are working at the level of triangulations and not at the level of geometries.
Hence, some triangulations may correspond to the same favourable Calabi-Yau geometry.
In \cite{Altman:2021pyc},
the 28,463 triangulation-wise involutions reduced to 5,660 geometry-wise proper involutions out of which 4,482 are obtained from favourable geometries.
In contrast,
the CICY orientifolds of \cite{Carta:2020ohw} are counted as distinct geometries.
In this sense,
the stated number of $\sim 7.2 \cdot 10^{7}$ should be taken with a grain of salt.}
The full data can be found in the following \href{https://github.com/AndreasSchachner/CY_Orientifold_database}{GitHub repository} together with a \texttt{jupyter} notebook providing instructions on how to read and work with the data.

The size of the D3-tadpole is critical for stabilising moduli with fluxes \cite{Dasgupta:1999ss,Giddings:2001yu}.
Recently, it has been argued that the required size of the D3-tadpole to stabilise all complex structure moduli with fluxes is $-Q_{\text{D3}}>\alpha\,  h^{1,2}$ \cite{Bena:2020xrh} where $\alpha>2/3$ in our convention.\footnote{In \cite{Bena:2020xrh}, $Q_{\text{D3}}=\chi(Y_{4})/24$, while in our convention $Q_{\text{D3}}=-\chi(Y_{4})/12$ as we compute the D3-charge in the perturbative type IIB double cover set up.} 
This is known as the tadpole problem since it is challenging to obtain such large D3 charges in typical type IIB orientifold models \cite{Bena:2020xrh,Braun:2020jrx}.\footnote{As pointed out e.g. in \cite{Gao:2022fdi},
the tadpole conjecture could be phrased more precisely by stating that the  landscape of vacua at large number of (complex structure) moduli may require singular geometries since the smoothness of the manifold was assumed to reach the conclusion.}

The cancellation of D7 tadpoles also plays a role in determining the size of the maximum possible D3 charge, as D7-branes and O7-planes induce some D3-charge. Usually this is done locally in terms of stacks of D7-branes on top of O7 orientifold planes. However, there are other means to cancel the tadpoles. In particular the consideration of Whitney branes, that cancel non-locally the D7-charge of the O7-planes , since they are not localised on top of the O7 planes, allows the possibility of substantially enhancing the maximum value of the D3 charge needed to cancel the D3 tadpoles.
We argue that construction with Whitney branes \cite{Collinucci:2008pf,Collinucci:2008sq} significantly surpass estimates for the D3-charge from $\mathrm{SO}(8)$ stacks of D7-branes on top of O7-planes.
Similar observations have been made in \cite{Carta:2020ohw} for general orientifolds of CICYs.
Our models beat previous records for the total D3-charges obtained in type IIB setups as exemplified by table~\ref{tab:OverviewD3ChargesLit}.
Ultimately, the goal is to combine our investigation with de Sitter constructions which we will explore in an upcoming paper \cite{Crino:2022P2}.

\begin{table}
\centering
\begin{tabular}{|c|c|c|c|c|}
\hline 
 &  &&  &  \\[-0.8em] 
$|Q_{\text{D3}}|$ & Type & D7-tadpole cancellation & $h^{1,1}$ & Reference \\ [0.3em]
\hline 
\hline 
 &  & & &  \\[-0.8em] 
 $\leq 428$ & KS & non-local & 3 &  \cite{Cicoli:2011qg} \\ [0.3em]
\hline 
 &  &&  &  \\[-0.8em] 
$\leq 72$ & CICY & local & $\leq 19$ &  \cite{Carta:2020ohw} \\ [0.3em]
\hline 
 &  & & &  \\[-0.8em] 
$\leq 264$ & CICY  & non-local & $\leq 19$ &  \cite{Carta:2020ohw} \\ [0.3em]
\hline 
 &  & & &  \\[-0.8em] 
$\leq 272$ & CICY & local & $4$ &  \cite{Cicoli:2021dhg} \\ [0.3em]
\hline 
 &  & & &  \\[-0.8em] 
$\leq 60$ & KS & local & $\leq 6$ &  \cite{Altman:2021pyc} \\ [0.3em]
\hline 
 &  & & &  \\[-0.8em] 
$\leq 504$ & KS & local & $\leq 12$ &  our database \\ [0.3em]
\hline 
 &  & & &  \\[-0.8em] 
$\leq 6664$ & KS & non-local & $\leq 12$ &  our database \\ [0.3em]
\hline 
\end{tabular} 
\caption{List of values for the total D3-charge contribution to the D3-tadpole.}\label{tab:OverviewD3ChargesLit} 
\end{table}

This paper is organised as follows. Next section is devoted to introductory material regarding the construction of CY orientifolds in terms of hypersurfaces of $4$-dimensional reflexive polytopes. We describe the different types of toric divisors and their topological properties that are relevant for our subsequent discussions. In section~\ref{sec:OrientifoldModels} we discuss the orientifold involution and determine the different brane configurations needed to cancel the tadpoles induced by the  O3 and O7 orientifold planes. In particular we point out the difference between local D7-branes and non-local D7 or Whitney branes and how they contribute differently to the D3 tadpoles. 

Section~\ref{sec:OrientifoldDatabase} describes in detail our database including the corresponding  Hodge numbers and  D3-brane charges, focusing on the general dependence of the D3 charges on the Hodge numbers and illustrating the maximum number of D3 charges that are relevant for the tadpole problem. First,
we present a full scan for orientifold models for $h^{1,1}\leq 7$. Then we  perform a random sampling for geometries with $8\leq h^{1,1}\leq 12$ and identify the largest values of D3 charges for both local and non-local D7-brane configurations.

We describe the model with the largest D3-charge contribution in our database explicitly in section~\ref{sec:H11_11Example}.
We summarise our conclusions in section~\ref{sec:Conclusions}. In appendix~\ref{app:EWB} we provide concrete examples of Whitney branes analysing their factorisation property depending on the topology of the divisors.
In a second appendix~\ref{app:GenusOneFib},
we present a simple example of a CY threefold with genus one fibration.

\section{From polytopes to Calabi-Yau hypersurfaces}

Here we collect some elementary definitions and formulae necessary for constructing CY hypersurfaces from $4$-dimensional reflexive polytopes in the Kreuzer-Skarke (KS) database \cite{Kreuzer:2000xy}, see \cite{Altman:2014bfa,Braun:2017nhi,Demirtas:2018akl} for details of the construction.

\subsection{Triangulations of 4D reflexive polyhedra}

We construct CY threefolds as anti-canonical hypersurfaces in $4$D Gorenstein toric Fano varieties \cite{Batyrev:1993oya}.
To this end, we use combinatorial information encoded in 4-dimensional reflexive lattice polytopes.
A complete list of 4D reflexive polytopes was initiated by Kreuzer and Skarke \cite{Kreuzer:2000xy}.
A database of CY threefolds with $h^{1,1}\leq 6$ was generated in \cite{Altman:2014bfa,Altman:2017vzk,Altman:2021pyc},
while \cite{Demirtas:2018akl,Demirtas:2020dbm,Demirtas:2021nlu,Demirtas:2021ote} explored regimes up to $h^{1,1}=491$ and their phenomenological implications in \cite{Mehta:2021pwf,Demirtas:2021gsq}.

To construct CY threefolds,
one begins with two reflexive polytopes $\Delta$ and $\Delta^{\circ}$ based on two $4$D lattices $M\cong \bZ^{4}$ and $N\cong \bZ^{4}$ with a pairing $\langle\cdot\, ,\cdot\rangle$ so that $\Delta\in M_{\bR}=M\otimes \bR$ and $\Delta^{\circ}\in N_{\bR}=N\otimes \bR$ satisfy
\begin{equation}
\langle\Delta ,\Delta^{\circ}\rangle\geq -1\, .
\end{equation}
We associate to the polytope $\Delta^{\circ}$ a fan $\Sigma$ in the following way. Reflexivity of $\Delta^{\circ}$ implies that the origin of $N$ is the unique interior lattice point of $\Delta^{\circ}$. We denote all other lattice points of $\Delta^{\circ}$ by $\nu_{i}$. The latter correspond to primitive generators of the rays of the fan $\Sigma$. The cones of $\Sigma$ are given by a triangulation of $\Delta^{\circ}$, i.e., special subsets of the $\nu_{i}$ with each containing the generators of a cone. We will focus on so-called \emph{fine, regular, star triangulations}\footnote{A triangulation is \emph{fine} if all points not interior to facets appear as vertices of a simplex.
Further, it is \emph{star} if the origin is a vertex of each full-dimensional simplex.
\emph{Regularity} implies that $\Sigma$ is the normal fan of a polytope and essentially ensures that $\bP_{\Sigma}$ and $X$ are projective, see \cite{de2010loera}.} (FRSTs), whose fan describes a simplicial toric $4$-fold denoted $\bP_{\Sigma}$. 
One can introduce weighted, homogeneous coordinates $z_{i}$ on $\bP_{\Sigma}$.
Within $\bP_{\Sigma}$, the CY threefold $X$ is found as the zero locus of a polynomial $P=\sum_m\, c_m\, p_m$, where $p_m$ are monomials in $z_i$'s and $c_m$ are coefficients related to the complex structure moduli of $X$. The individual monomials $p_{m}$ appearing in $P$ are encoded by $\Delta$, also called the \emph{Newton polytope} of the hypersurface.
They are easily computed from (see e.g. Eq.~(A.8) in \cite{Altman:2014bfa})
\begin{equation}\label{eq:CYMonomials} 
p_{m}=\prod_{i}\, z_{i}^{\langle m,\nu_{i}\rangle+1}\kom m\in \Delta\cap M\:.
\end{equation}

Although $\bP_{\Sigma}$ does not need to be smooth, every FRST leads to a smooth hypersurface $X$ \cite{Batyrev:1993oya}.
We focus exclusively on \emph{favourable} geometries where
\begin{equation}
h^{1,1}(X)=
\mathrm{dim}(\mathrm{Pic}(\bP_{\Sigma})) \:,
\end{equation}
that is, the Kähler moduli on $X$ descend from those of the ambient space $\bP_{\Sigma}$.

Computationally, it is generically expensive to compute all triangulations for a given $\Delta^{\circ}$. For sufficiently simple polytopes, that is, those with few lattice points, all triangulations were obtained in \cite{Altman:2014bfa} up to $h^{1,1}(X)=6$. Here, only a small subset of the triangulation data was required to define the geometry of $X$. Specifically, everything happening inside faces of co-dimension one can be ignored.
In our scan, we check all favourable geometries for $h^{1,1}(X)\leq 7$ and provide partial results up to $h^{1,1}(X)=12$.

\subsection{Toric divisors and their topologies}\label{sec:DivTops}

Each weighted, homogeneous coordinate $z_{i}$ of $\bP_\Sigma$
corresponds to a point on the boundary of $\Delta^{\circ}$. The loci $\tilde D_i=\lbrace z_{i}=0\rbrace$ are called \emph{prime toric divisors} (see e.g. \cite{Demirtas:2018akl} for details).
The subset of such divisors which intersect $X$ transversely corresponds to points that lie in faces of $\Delta^{\circ}$ of dimension $\leq 2$. 
Intersecting such a locus with the CY hypersurface equation, one gets a divisor 
$D_i\in H^{1,1}(X,\mathbb{Z})$ which defines a $4$-cycle in $X$ dual to a $2$-cycle $\omega_{i}$.
Since we focus exclusively on favourable polytopes and geometries,
all such prime toric divisors are irreducible on $X$.
Hence,
$H_{4}(X,\bZ)$ is generated by any basis constructed from $\lbrace D_{i}\rbrace$, $i=1,\ldots , h^{1,1}(X)+4$.

The Hodge numbers of divisors are collectively denoted as
\begin{equation}
h^{\bullet}(D)=\lbrace h^{0,0}(D), h^{0,1}(D), h^{0,2}(D), h^{1,1}(D)\rbrace\, .
\end{equation}
A rigid divisor $D_{\text{rig}}$ is defined as
\begin{equation}
h^{\bullet}(D_{\text{rig}})=\lbrace 1,0,0, h^{1,1}(D_{\text{rig}})\rbrace\, .
\end{equation}
Prototypical examples include del Pezzo divisors  $\text{dP}_{n}$, $n=0,\ldots ,8$ (where $\text{dP}_{0}=\mathbb{P}^{2}$) and the Hirzebruch surface $\mathbb{F}_{0}=\bP^{1}\times \bP^{1}$, for which $h^{(1,1)}(D_{\text{dP}_{n}})=n+1$ and $h^{(1,1)}(D_{\mathbb{F}_{0}})=2$.
These divisors play a special role since they can be shrunk to a point allowing for SM realisations on D3-branes placed at the tip of the singularity \cite{Cicoli:2021dhg}.
Rigid divisors with $h^{(1,1)}(D)>9$ are typically referred to as \emph{non-shrinkable}.

For later purposes,
we distinguish other common types of divisors as follows, see also \cite{Altman:2021pyc}:
\begin{enumerate}
\item Wilson divisors: $h^{\bullet}(D)=\lbrace 1,h^{1,0},0,h^{1,1}\rbrace$ with both $h^{1,0}\neq 0$ and $h^{1,1}\neq 0$,
\item K3 divisor: $h^{\bullet}(D)=\lbrace 1,0,1,20\rbrace$,
\item SD1: $h^{\bullet}(D)=\lbrace 1,0,1,21\rbrace$,
\item SD2: $h^{\bullet}(D)=\lbrace 1,0,2,30\rbrace$.
\end{enumerate}

To compute these Hodge numbers,
we follow the steps outlined in \cite{Braun:2015pza,Braun:2017nhi}, that we now review.\footnote{Another way to computing divisor topologies uses the \texttt{cohomCalg} package \cite{Blumenhagen:2010pv, Blumenhagen:2011xn} which is however limited when applied to models with $h^{1,1}(X)\geq 6$. In particular, the authors of \cite{Altman:2021pyc} computed the Hodge numbers of divisors up to $h^{1,1}(X)=6$ in this way.}
As said above, each toric divisor $D_{i}\in H^{1,1}(X,\mathbb{Z})$  is associated with a lattice point $\nu_{i}$ on $\Delta^{\circ}$.
Its Hodge numbers $h^{0,p}$ can be obtained from the location of $\nu_{i}$ inside $\Delta^{\circ}$.
In fact, one finds the following \cite{Batyrev:1993oya,danilov1987newton}:
\begin{enumerate}
\item \emph{Rigid divisors:} A toric divisor $D_{i}$ is rigid if
\begin{equation}
\ell^{*}(\Theta)=0\:,
\end{equation}
where $\ell^{*}$ is the sum of all interior points of the face $\Theta$, which is the dual of the face containing $\nu_{i}$.
\item \emph{Deformation divisors:} Divisors with $h^{0,2}(D_{i})>0$ and $h^{0,1}(D_{i})=0$ are associated with points $\nu_{i}$ corresponding to vertices of $\Delta^{\circ}$ so that
\begin{equation}
h^{0,1}(D_{i})=0\kom h^{0,2}(D_{i})=\ell^{*}(\Theta^{[3]})\:,
\end{equation}
in terms of the dual face $\Theta^{[3]}$ to $\nu_{i}=\Theta^{\circ[0]}$.
\item \emph{Wilson divisors:} Lastly,
divisors $D_{i}$ associated with points $\nu_{i}$ inside a one-dimensional face $\Theta^{\circ[1]}$ of $\Delta^{\circ}$ give rise to
\begin{equation}
h^{0,1}(D_{i})=\ell^{*}(\Theta^{[2]})\kom h^{0,2}(D_{i})=0\:,
\end{equation}
in terms of the dual face $\Theta^{[2]}$ to $\Theta^{\circ[1]}$.
\end{enumerate}
The above conditions can easily be checked using \href{https://www.sagemath.org}{\texttt{Sage}} \cite{sagemath}.
The remaining Hodge numbers can then be inferred from the Euler characteristic and the arithmetic genus
\begin{align}
\label{eq:EulerDiv} \chi(D)&=2h^{0,0}-4h^{0,1}+2h^{0,2}+h^{1,1}=\int_D c_2(D)\, ,\\
\label{eq:AGDiv} \chi_{0}(D)&=h^{0,0}-h^{0,1}+h^{0,2}=\dfrac{1}{12}\int_D\left (c_1(D)^2+c_2(D) \right)\, .
\end{align}
The RHS can be easily computed from the CY data, by using adjunction formula $
c_{2}(X)=c_{2}(D)-c_{1}(D)^{2}$ and $c_{1}(D)=-\iota^*D$ for a CY:
\begin{equation}
\int_D c_2(D)= \int_{D}\, (D^{2}+c_{2}(X)) \kom \int_D\left (c_1(D)^2+c_2(D) \right)=\int_{D}\, (2D^{2}+c_{2}(X))\:.
\end{equation}
The above can be solved for $h^{0,0}$ and $h^{1,1}$ as
\begin{align}
h^{0,0}=\chi_{0}(D)+h^{0,1}-h^{0,2}\kom h^{1,1}=\chi(D)-2\chi_{0}(D)+2h^{0,1}\, .
\end{align}

Instead of computing Hodge numbers explicitly,
it can also be useful to check the sufficient conditions for del Pezzo divisors using their intersection numbers.
Indeed,
a del Pezzo divisor must satisfy the following topological conditions
\begin{equation}\label{eq:dP}
\int_{X} D_s^3 = k_{sss}=9 - n > 0\, , \qquad \int_{X} D_s^2 \, D_i \leq 0 \qquad \forall \, i \neq s \,.
\end{equation}
We moreover look for divisors $D_s$ that satisfy the following \emph{diagonality} condition \cite{Cicoli:2018tcq}
\begin{equation}\label{eq:diagdP}
k_{sss} \, \, k_{s i j } = k_{ss i} \, \, k_{ss j} \, \qquad \qquad \forall \, \, \, i, j \:.
\end{equation}
If this condition is satisfied, then the volume of the associated 4-cycle $D_s$ is a complete-square:
\begin{equation}
 \tau_s = \frac{1}{2}\, k_{s ij} t^i \, t^j = \frac{1}{2 \, k_{sss}}\, k_{ssi} \, k_{s s j} t^i \, t^j = \frac{1}{2 \, k_{sss}}\, \left(k_{ss i} \,t^i \, \right)^2\, ,
\end{equation}
where we sum over $i,j$ but not over $s$.
This condition is commonly used in the LVS \cite{Balasubramanian:2005zx,Conlon:2005ki} by ensuring that the volume form is of swiss cheese type.
Furthermore,
it allows to generate del Pezzo singularities by shrinking the divisor to a point along one direction of the K\"ahler moduli space which is heavily utilised in constructions of branes at singularities, see \cite{Cicoli:2021dhg} for a recent discussion and further references.

\section{Orientifold configurations}\label{sec:OrientifoldModels}

We focus on involutions of toric coordinates of the form $\sigma_{k}:\, z_{k}\raw -z_{k}$ for which $h^{1,1}_{-}=0$ (if the corresponding geometry is favourable, see e.g. \cite{Altman:2021pyc} for a discussion).
For each involution,
we obtain configurations of O$p$-planes given by fixed point loci of the associated involution $\sigma_{k}$.
Tadpole and anomaly cancellation is ensured by adding an appropriate D-brane setup.

\subsection{Orientifold data}\label{sec:OrientifoldData} 

We consider involutions with O3/O7 orientifold planes. An O7-plane wraps a fixed surface $D_i$ in the CY three-fold, while an O3-plane is at an isolated fixed point of the involution.

An important topological invariant that we will need later to compute the D3-charge contributions is the Euler characteristic \eqref{eq:EulerDiv} of a (smooth) divisors $D_{i}$. As said above, it is given by the integral $\int_{D_{i}}\, c_{2}(D_{i})$ which is computed from the topological data
\begin{align}
\label{eq:ChiDivisorTripNums} 
\chi(D_{i})&=\kappa_{iii}+\int_{D_{i}}\, c_{2}(X)\\
\label{eq:ChiDivisorHodgeNums}
&= 2h^{0,0}(D_{i})-4h^{1,0}(D_{i})+2h^{2,0}(D_{i})+h^{1,1}(D_{i})\, .
\end{align}

The knowledge of the topology of the fixed point set allows to compute other integers invariants of the CY orientifold. In particular the cohomology groups $H^{p,q}(X)$ split into even and odd subspaces of the (pull-back of the) orientifold involution. Their dimensions are called $h^{p,1}_+(X)$ and $h^{p,1}_-(X)$ respectively. 
To compute them,
we use Lefschetz fixed point theorem which states that
\begin{equation}\label{LefTh}
\sum_{i}\, (-1)^{i}(b_{+}^{i}(X)-b_{-}^{i}(X))=\chi(O_{\sigma})\kom b_{\pm}^{i}(X)=\sum_{p+q=i}\, h_{\pm}^{p,q}(X)\, 
\end{equation}
in terms of the even/odd Betti numbers $b_{\pm}^{i}(X)$.
Here, we will have
\begin{equation}\label{chiOsigmaO7O3}
\chi(O_{\sigma})=\sum_{i}\,  
\chi(O7_{i})+\sum_{k}\, 
\chi(O3_{k})\, , \qquad\mbox{with }\chi(O3_k)=1\:.
\end{equation}
For CY threefolds, the expression \eqref{LefTh} simplifies to
\begin{equation}\label{eq:HodgeNumOrientCond} 
2+2\left (h^{1,1}_{+}(X)-h^{1,1}_{-}(X)\right )-2\left (h^{1,2}_{+}(X)-h^{1,2}_{-}(X)-1\right )=\chi(O_{\sigma})\, .
\end{equation}
Since we know $h^{1,1}_\pm(X)$ (in cases under study $h^{1,1}_-(X)=0$ and $h^{1,1}_+(X)=h^{1,1}(X)$), we can use this relation to obtain the Hodge numbers $h^{1,2}_{\pm}(X)$. We need to solve the equations:
\begin{align}\label{eq:HodgeNumbersH12Orientifold} 
h^{1,2}_{+}(X)+h^{1,2}_{-}(X)=h^{1,2}(X)\kom h^{1,2}_{+}(X)-h^{1,2}_{-}(X)=h^{1,1}(X)+2-\dfrac{\chi(O_{\sigma})}{2}\, .
\end{align}
Below, we use this data to discard models where the computation of $h^{1,2}_{\pm}(X)$ lead to non-integer values, as this is a signal of possible unwanted singularities.

To detect more subtle singularities which are not manifest in the orientifold data,
we look at the underlying polytopes.
Let us just reiterate again that we are interested in involutions of a single homogeneous toric coordinate $z_{k}\raw -z_{k}$ which is associated with one of the boundary lattice points $\nu_{k}$ of $\Delta^{\circ}$ not interior to facets (i.e., 3-faces).
Recalling \eqref{eq:CYMonomials},
the invariant CY equation for $z_{k}\raw -z_{k}$ is obtained from the monomials
\begin{equation} 
p_{m}^{(k)}=\prod_{i}\, z_{i}^{\langle m,\nu_{i}\rangle+1}\kom m\in \Delta_{k}\cap M
\end{equation}
where we define
\begin{equation}
\Delta_{k}=\lbrace m\in \Delta \, :\, \langle m,\nu_{k}\rangle+1\in 2\bN\rbrace\, .
\end{equation}
We argue that the properties of $\Delta_{k}$ are in one-to-one correspondence with the hypersurface obtained from tuning the CY equation to be invariant under $z_{k}\raw -z_{k}$.

Removing (the non-invariant) monomials for the CY defining equation can force some singularities: either 1)  the hypersurface is forced to touch singularities of the ambient space, or 2) the defining polynomial describes now a singular hypersurface (there are points where the differential of the equation and the equation itself vanish simultaneously). Since we want to work with smooth spaces, we need to discard models where the involution forces singularities.

The desired invariant CY $X$ can be obtained from triangulations of the polar dual $\Delta_{k}^{\circ}$. 
Since we are interested in collecting big numbers, we decide to keep in the analysis only invariant CY's corresponding to favorable $\Delta_{k}^{\circ}$ and reflexive $\Delta_{k}$. For these CY we can claim smoothness. We checked in several models that the excluded CY's were actually singular.\footnote{Of course, string theory is well defined also on singular spaces. In the database we provide on \href{https://github.com/AndreasSchachner/CY_Orientifold_database}{GitHub},
the reader can find also the data of the singular models. However we decide to stay on the safe side, studying models with smooth geometry where the usual formulae to compute topological invariants work well.}

\subsection{D7-branes}

In order to cancel the D7-tadpole induced by the O7-planes,
we add D7-branes on the appropriate divisors. 

The D7-charge of an O7-plane wrapping the divisor $D_i$ is $-8[D_{i}]$.
The easiest way to cancel the D7-tadpole is then to put 4 D7-branes plus their 4 images on top of the O7-plane. The D7-brane configuration is given in this case by $z_i^8=0$. The gauge group supported on such a stack is $\mathrm{SO}(8)$.

The other extreme case is to cancel the D7-tadpole by a fully recombined D7-brane in the homology class $8[D_{i}]$. 
This is called \emph{Whitney brane}, as it is forced to have a singular worldvolume of the form of the Whitney umbrella \cite{Collinucci:2008pf,Braun:2008ua,Collinucci:2008sq}:
\begin{equation}\label{eq:DefEqWhitneyBrane}
\eta^{2}-z_i^{2}\chi=0\:,
\end{equation}
where $z_i\in\mathcal{O}(D_{i})$, $\eta\in\mathcal{O}(4D_{i})$ and $\chi\in\mathcal{O}(6D_{i})$. The sections $\eta$ and $\chi$ are invariant under the orientifold involution, while $z_i\mapsto-z_i$.
This brane supports no continuous gauge group 
and has zero chiral intersection with (fluxless) D7- or E3-branes supported on an intersecting divisor \cite{Collinucci:2008sq}.

For a generic toric divisor with high weights, the locus \eqref{eq:DefEqWhitneyBrane} is connected. However, there can be particular cases when the generic sections $\eta,\chi$ of the line bundles $\mathcal{O}(4D_i),\mathcal{O}(6D_i)$ factorise. For instance, it may happen that 
\begin{equation}\label{eq:WhitFactor}
\eta=z_{j}^m\eta'\kom \chi=z_{j}^{2m}\chi'\, .
\end{equation}
Then the equation of the configuration will be
\begin{equation}
z_{j}^{2m}(\eta'^2-z_{i}^2\chi')=0\, .
\end{equation}
If this happens, we recover a stack of D7-branes on $z_{j}=0$ plus a Whitney brane of lower degree in the homology class $8[D_{i}]-2m[D_{j}]$ (see e.g. \cite{Cicoli:2011qg}). 

A particular important example of a factorisation like \eqref{eq:WhitFactor} appears when $D_i$ is a \emph{rigid divisor}. In this case $\eta \propto z_i^4$ and $\chi\propto z_i^6$ and we are left with a configuration, whose locus is $z_i^8=0$, i.e. we have four D7-branes plus their four images on top of the O7-plane, generating an $\mathrm{SO}(8)$ D7-brane stack.

Also special non-rigid divisors can lead to a factorisation of the Whitney brane.
In fact, whenever $D_{i}$ is a K3 surface the Whitney brane splits into a $\mathrm{U}(1)^4$ configuration with four D7-branes plus their four images.
In our analysis we found that this kind of factorization often happens for divisors with $h^{2,0}=1$. 
We provide two explicit examples in App.~\ref{app:EWB}.

\subsection{D-brane worldvolume flux}\label{sec:WorldVolFlux}

Let us assume that the orientifold model contains stacks of E3/D7-branes wrapped on a divisors $D$.
We can then turn on a gauge flux
\begin{equation}
\cF=F_{2}-\iota^{*}B_{2}\:,
\end{equation}
where $F_2$ is the field strength of the worldvolume U(1) gauge theory, $B_2$ is the NSNS 2-form potential and  $\iota^{*}:\, H^2(X)\raw H^2(D)$ is the pull-back map on $D$.

Freed-Witten anomaly cancellation \cite{Freed:1999vc} requires the following quantization condition on $F_2$:
\begin{equation}
F_{2}+\dfrac{c_{1}(D)}{2}\in H^{2}(D,\mathbb{Z})\:,
\end{equation}
where $c_{1}(D)=-\iota^{*}D$ for $X$ a CY.
This implies that the following expression for $\cF$ fulfills this condition:
\begin{equation}\label{eq:FluxGenDef} 
\cF=\sum_{k=1}^{h^{1,1}(X)} n_k\,\iota^{*}D_{k} + \dfrac{1}{2}\iota^{*}D-\iota^{*}B_{2}\qquad\mbox{with}\qquad n_k\in\mathbb{Z}\:
\end{equation}
and with $\{D_k\}$ an integral basis of $H^2(X,\mathbb{Z})$.

If $D$ is wrapped by an O(1) ED3-instanton, then the orientifold invariance of the configuration requires 
\begin{equation}
\cF_{\text{ED3}}=0\, .
\end{equation}
This can be achieved by properly choosing the background of $B_2$, i.e. s.t.
\begin{equation}\label{B2E3inst}
\iota^{*}B_{2}=\iota^{*}D_\mathbb{Z} \,-\dfrac{1}{2}\iota^{*}D\,\qquad\mbox{with }\, D_\mathbb{Z}\in H^2(X,\mathbb{Z})\:.
\end{equation}
Rank-2 E3 instantons with a non-trivial gauge bundle can also be allowed by a $B_2$ that does not fulfill \eqref{B2E3inst} \cite{Berglund:2012gr}.

\

Let us come to the Whitney brane \eqref{eq:DefEqWhitneyBrane} in the homology class $8[D_i]$. The Whitney brane can support an integral flux, that as we will see contribute to the D3-charge. When this flux is present, the defining polynomial $\chi$ is forced to take the form $\chi=\psi^2-\rho\tau$ with $\psi\in\mathcal{O}(3D_i)$, $\rho\in \mathcal{O}(3D_i-2F+2B_2)$ and $\tau\in\mathcal{O}(3D_i+2F-2B_2)$, where $F\in H^2(X,\mathbb{Z})$ and $B_2$ is the B-field \cite{Collinucci:2008pf}. The flux data is encoded in the choice of the line bundles
\begin{equation}\label{LineBndlsFluxWhit}
 \mathcal{O}(3D_i-2F+2B_2)\qquad\mbox{ and }\qquad\mathcal{O}(3D_i+2F-2B_2),
\end{equation}
i.e. in the choice of the integral two-form $F$. One gets a zero flux when one of these line bundles is trivial. Notice that this can be achieved when $D_i+2B$ is an even form (remember that $B$ can take half-integral values), that may not happen.

When one of the line bundles in \eqref{LineBndlsFluxWhit} has no holomorphic sections, then either $\rho$ or $\tau$ is forced to vanish.
In this case, the Whitney brane locus factorizes as
\begin{equation}\label{WhitIntoD7D7p}
   (\eta + z_i \psi)(\eta - z_i \psi)=0\:,
\end{equation}
i.e. it splits into a pair of one D7-brane and its orientifold image, both in the homology class $[4D_i]$.
Hence, in order for the Whitney brane to be non-factorised one requires that the line bundles \eqref{LineBndlsFluxWhit} have holomorphic sections, i.e. 
\begin{equation}\label{CondFluxFWhit}
-\frac{3D_{i}}{2} + B_2 \leq F\leq \frac{3D_{i}}{2} + B_2 \:.
\end{equation}

Even when the condition \eqref{CondFluxFWhit} holds, one can set $\rho=\tau=0$ by a deformation of the Whitney brane. Correspondingly, the Whitney branes splits as in \eqref{WhitIntoD7D7p}. The $\mathrm{U}(1)$ D7-brane has then a flux $\mathcal{F}=\iota^*(F-B_2)$, where $F$ is the same two-form appearing in \eqref{LineBndlsFluxWhit}. 
When this happens, the D7-brane can have chiral intersections with some E3-instantons. This will be counted by the formula
\begin{equation}\label{eq:FluxConstraintD3D7} 
0=\int_{\text{D7}\cap \text{ED3}}\, (\cF_{\text{D7}}-\cF_{\text{ED3}})=D_{\text{D7}}\cdot D_{\text{ED3}}\cdot \cF_{\text{D7}}\, .
\end{equation}
A non-perturbative instanton contribution to the $4$D superpotential requires the absence of chiral modes (for non-chiral modes, see footnote 21 in \cite{Cicoli:2021dhg}) at the intersection of D7-branes and ED3-instantons.
This generally limits the flux allowed on the D7-branes.

A $\mathrm{U}(1)$ D7-brane with fluxes supports a generically non-zero FI-term:
\begin{equation}
\xi_{\text{D7}}=\dfrac{1}{4\pi\cV}\int_{D_{i}}\, \cF_{\text{D7}}\wedge J\, .
\end{equation}
This term, if non-zero, 
requires a non-vanishing VEV for scalar modes at the intersection between D7 and its image in order to cancel the D-term potential.\footnote{There is also the possibility that the sign of $\xi$ generates a D-term that is strictly positive; in this case this leads to SUSY breaking \cite{Collinucci:2014qfa}.} If $F$ satisfies \eqref{CondFluxFWhit}, this corresponds to deforming the branes switching on non-zero $\rho$ and $\tau$; this recombines the two branes into a Whitney brane. If the condition \eqref{CondFluxFWhit} is not fulfilled, then only $\rho$ or $\tau$ can be non-zero and we generate a T-brane background, i.e. the two branes form a bound state whose locus is still \eqref{WhitIntoD7D7p} \cite{Collinucci:2014qfa}.

\subsection{The D3-tadpole}\label{sec:D3Tadpole}

We now compute the induced D3-charge from the orientifold configuration.
Generally, the D3-tadpole cancellation condition reads
\begin{equation}\label{eq:DThreeTadpoleGen} 
N_{\text{D}3}+N_{\text{D}3^{\prime}}+N_{\text{flux}}=-Q_{\text{D}3}\kom \mbox{ with }Q_{\text{D}3}=Q_{\text{D}7}^{\text{tot}}+Q_{Op}^{\text{tot}}\:,
\end{equation}
where
\begin{align}
Q_{\text{D}7}^{\text{tot}}=\sum_{i}\, (Q^{i}_{\text{D}7}+Q^{i}_{D7^{\prime}})\kom Q_{Op}^{\text{tot}}=\sum_{k}\, Q_{\text{O}3}^{k}+\sum_{l}\, Q_{\text{O}7}^{l}\, .
\end{align}

For O7/O3-planes,
we collect
\begin{equation}
Q_{\text{O}7}^{i}=-\dfrac{\chi(D_{i})}{6}\kom Q_{\text{O}3}^{i}=-\dfrac{1}{2}\, ,
\end{equation}
whereas a $\mathrm{U}(1)$ D7-brane contributes as
\begin{equation}
Q_{\text{D}7}^{i}=-\dfrac{\chi(D_{i})}{24}-\dfrac{1}{2}\int_{D_{i}}\cF\wedge\cF\, .
\end{equation}

For later convenience,
we refer to $Q_{\text{SO}(8)}^{\text{tot}}$ and $Q_{\text{WD7}}^{\text{tot}}$ as the D3-charge contribution coming from rigid D7-branes and Whitney branes respectively.
The first one is easy to compute: when all the four D7 branes have the same flux $\cal F$, then the group is broken to $\mathrm{SU}(4)$ (the diagonal $\mathrm{U}(1)$ get a St\"uckelberg mass due to Green-Schwartz mechanism) and the contribution of the stack to the D3-tadpole is
\begin{equation}\label{D3chSO8}
 Q_{\text{SO}(8)}^{\text{tot}} = 8Q_{\rm one\, D7} = -\dfrac{\chi(D_{i})}{3}-4\int_{D_{i}}\cF\wedge\cF\:.
\end{equation}

For the Whitney brane the situation is a bit different. The expression of its total D3-charge can be derived in a simple way \cite{Collinucci:2008sq}: the D3-charge does not change under recombination or splitting of branes; hence we can compute it in the phase where the Whitney brane splits into a $\mathrm{U}(1)$ brane and its image. Hence, for a Whitney brane in the class $8D_i$
\begin{equation}\label{D3chargeWhitBr}
Q_{WD7}^{i}=-\dfrac{\chi(4D_{i})}{12}-\int_{X}D_{i}\wedge(F-B_2)\wedge(F-B_2)\, .
\end{equation}
with $F$ given in \eqref{CondFluxFWhit}. The geometric contribution of the Whitney brane is different from the geometric contribution of the brane/image-brane system  \cite{Collinucci:2008pf}. In fact, the D3-charge contributions from geometry and from the flux encoded into the line bundles \eqref{LineBndlsFluxWhit} are \cite{Collinucci:2008pf,Braun:2011zm} 
\begin{eqnarray}
\label{D3WD7geom}Q_{WD7,{\rm geom}}^{i} &=&  -  \frac{1}{3}\int_{X} D_i\wedge \left( 43 D_i\wedge D_i + c_2(X) \right) = -\frac{\chi(4D_i)}{12} -9\int_X D_i^3\:, \\
\label{D3WD7flux}Q_{WD7,{\rm flux}}^{i} &=& \int_{X} D_i \wedge (3D_i-2F+2B_2)\wedge (3D_i + 2F - 2B_2)\:.
\end{eqnarray}
One can easily check that the sum of the two gives \eqref{D3chargeWhitBr} and that $Q_{WD7,{\rm flux}}^{i}$ is identically zero when the line bundles \eqref{LineBndlsFluxWhit} are trivial. If $D_i+2B_2$ is an even integral form, one can actually take zero flux and make $Q_{WD7,{\rm flux}}^{i} $ vanish.\footnote{Actually, it is enough that $\iota_{D_i}^* (3D_i+2B_2-2F)$ vanishes, in order to have $Q_{WD7,{\rm flux}}^{i}=0$.} Generically this is not possible, but it is always possible to choose $F$ such that $Q_{WD7,{\rm flux}}^{i} \ll Q_{WD7,{\rm geom}}^{i}$. This will justify, in our analysis, to approximate the D3-charge of a Whitney brane by its geometric contribution.

\

Before we continue,
let us make a few estimates on the D3-charge contributions.
Let us start from
\begin{equation}\label{eq:D3OpHodgeNum}
-Q_{\text{O}7}^{\text{tot}}= \sum_{i}\, \dfrac{\chi(D_{i})}{6}= \dfrac{\chi(O_{\sigma})-N_{\text{O}3}}{6}=\dfrac{h^{1,1}(X)+h^{1,2}_{-}(X)-h^{1,2}_{+}(X)+2}{3}-\frac{N_{\text{O}3}}{6}\, ,\nonumber
\end{equation}
where we used \eqref{chiOsigmaO7O3} and \eqref{eq:HodgeNumOrientCond} (with $h^{1,1}_-(X)=0$) and where $N_{\text{O}3}$ is the number of isolated fixed points in $X$. 
We conclude that there are two possibilities of increasing this value by investigating models with either many Kähler or instead many complex structure moduli.
We are going to observe this scaling with respect to $h^{1,2}_{-}$ quite frequently below for orientifolds with $h^{1,2}_{+}=0$, see in particular Fig.~\ref{fig:OverviewD3ChargeOp}.

Now, assume we have O7-planes on divisors $D_i$ and that we cancel their D7-charge by putting 4 D7-branes plus their 4 images in each $D_i$ (producing a bunch of $\mathrm{SO}(8)$ stacks). This is the choice that minimize the D3-charge contribution from O7/D7's. 
Let us now consider several CY's $X$ and involutions and let us estimate what is the maximum that we can get for the D3-charge for this minimal configuration, where we cancel the D7-tadpole locally (i.e. with only $\mathrm{SO}(8)$ stacks). 
One may use \eqref{D3chSO8} and write (in the absence of worldvolume fluxes)
\begin{equation}
-Q_{\text{SO}(8)}^{\text{tot}}=\sum_{i}\dfrac{\chi(D_{i})}{3}= 2\dfrac{h^{1,1}(X)+h^{1,2}_{-}(X)-h^{1,2}_{+}(X)+2}{3}-\frac{N_{\text{O}3}}{3}
\end{equation}
to arrive at \cite{Collinucci:2008pf,Carta:2020ohw,Bena:2020xrh,Gao:2022fdi}
\begin{equation}\label{eq:BoundD3ChargeLocD7} 
-Q_{\text{D3}} = -Q_{\text{O}3}^{\text{tot}}  -Q_{\text{O}7}^{\text{tot}} - Q_{\text{SO}(8)}^{\text{tot}} 
  \leq 2+h^{1,1}+h^{1,2}\, ,
\end{equation}
where in the last step we used $Q_{\text{O}3}^{\text{tot}} = - \frac{N_{\text{O}3}}{2}$ and the fact that $h^{1,2}_{-}(X)-h^{1,2}_{+}(X)\leq h^{1,2}(X)$.

In the KS database, this implies $-Q_{\text{D3}}\leq 504$ for e.g. CYs with Hodge numbers $(h^{1,1},h^{1,2})=(11,491)$ which we discuss further below.

\subsubsection*{D3-tadpole in F-theory}

A perturbative type IIB orientifold compactification can always be described in F-theory language. The F-theory compactification manifold is a CY fourfold that is an elliptic fibration over the base space $B_3=X/\sigma$, that is the quotient of $X$ by the orientifold involution. 
If the involution allows to cancel all the D7-tadpoles by Whitney branes, this corresponds to a smooth CY fourfold in F-theory. Splitting the Whitney branes in type IIB, producing a non-trivial gauge group $G$, corresponds to deforming the fourfold  generating codimension-3 (abelian $G$) or codimension-2 (non-abelian $G$) singularities. If the fixed point locus includes a rigid divisor, then the D7-branes on that divisor support an $\mathrm{SO}(8)$ gauge group that cannot be deformed; this corresponds to a so called non-Higgsable cluster in the F-theory fourfold \cite{Morrison:2012np,Morrison:2014lca}, i.e. in this case a non-deformable $D_4$ singularity.

The D3-tadpole cancellation condition in F-theory takes the form:\footnote{We note that $\chi(Y_{4})/24=-Q_{\text{D3}}/2$ when compared to D3-tadpole in \eqref{eq:DThreeTadpoleGen} given that we work with the double cover in Sect.~\ref{sec:D3Tadpole}.}
\begin{equation}
\dfrac{1}{2}\int_{Y_{4}}\, G_{4}\wedge G_{4}+N_{\text{D3}}=\dfrac{\chi(Y_{4})}{24}\:,
\end{equation}
where\footnote{This is obtained from $\chi(Y_{4})=4+2h^{1,1}-4h^{1,2}+2h^{1,3}+h^{2,2}$ together with $h^{2,2}=44+4h^{1,1}-2h^{1,2}+4h^{1,3}$.}
$
\chi(Y_{4})=6\left (8+h^{1,1}(Y_{4})+h^{1,3}(Y_{4})-h^{1,2}(Y_{4})\right )
$
is the Euler characteristic of the fourfold. When the fourfold is singular, this formula still applies, provided one uses the resolved fourfold \cite{Collinucci:2008pf, Grimm:2010ez,Braun:2011zm}. However, the geometric contribution to the tadpole decreases as one makes a deformation from a smooth to a singular fourfold (with some gauge group and matter). This is consistent with what one observes in type IIB: splitting the Whitney brane, the D3 contribution decreases (in absolute value) \cite{Collinucci:2008pf}.

The large D3-charges that are usually mentioned in literature as coming from F-theory backgrounds, correspond typically to smooth fourfold (with no gauge group or matter). These large D3-charges can be reached in type IIB by canceling the D7-tadpole by means of Whitney branes.

\section{Orientifold database}\label{sec:OrientifoldDatabase}

In this section,
we generate a database of CY orientifolds models based on the general information summarised in Sect.~\ref{sec:OrientifoldModels}.
An essential tool in this context is the \texttt{CYTools} package \cite{Demirtas:2020dbm} which allows us to construct FRSTs from polytopes at arbitrary $h^{1,1}$.
Beyond that,
we implemented a basic algorithm to construct CY orientifolds from the polytope and triangulation data from reflection involutions.
We test this implementation up to $h^{1,1}= 12$.
As an application of our database,
we investigate the size of D3-charge contributions.

\subsection{An algorithm for finding orientifold configurations}

For each CY $X$ and each choice of involution, we determine the fixed point set in the following way.
\begin{enumerate}
\item We first find the CY equation that is symmetric under the chosen involution, by 
determining the set of invariant monomials under $\sigma_{k}$ (keeping only those in \eqref{eq:CYMonomials} involving even powers of $z_{k}$).
\item We determine loci of points of the toric ambient fourfold $\mathbb{P}_\Sigma$ that are fixed under $\sigma_{k}$: in practice, we consider the action on the coordinates $z_{j}$ of $\sigma_{k}$ and $\sigma_{k}\cdot \zeta_a$, with $\zeta_a$ the $\mathbb{C}^*$ toric equivalences, and taking into account the SR ideal.
\item We check whether the invariant CY equation vanishes at a given locus.
If no, a complex co-dimension $n$ locus in $\mathbb{P}_\Sigma$ determines the presence of an O$m$-plane with $m=3+2(3-n)$.
If yes,
a co-dimension $n$ locus corresponds to an O$m$-plane with $m=3+2(4-n)$.
\end{enumerate}
We consider involutions that generate O3- and O7-planes, so in our scan there are no O5/O9-planes which can however arise for exchange involutions \cite{Altman:2021pyc}.

The number of O3-planes is determined from the intersection numbers either in the CY for co-dimension $3$ or in the ambient fourfold for co-dimension $4$ fixed point loci.
The latter can be obtained from \texttt{CYTools} where we take special care of singularities in the ambient space.

A similar algorithm to determine the O-plane configurations in the context of exchange involutions was introduced in \cite{Altman:2021pyc}.
In this sense,
our work provides a complementary analysis for the geometries with $h^{1,1}\leq 6$,
while providing additional statistics up to $h^{1,1}=12$.
What sets our database apart is the study of Whitney brane configurations as opposed to simple $\mathrm{SO}(8)$ stacks of D7-branes.

\subsection{Complete scan for CYs with $h^{1,1}\leq 7$ and random CYs at $h^{1,1}\leq 12$}

\begin{table}[t!]
\centering
\resizebox{\columnwidth}{!}{
\begin{tabular}{|c|c|c|c|c|c|c||c|}
\hline 
&  &  &  &  &  &  & \\ [-0.8em]
$h^{1,1}$ & 2 & 3 & 4 & 5 & 6 & 7  & total \\ [0.3em]
\hline 
\hline 
 &  &  &  &  &  &  & \\ [-0.8em]
polytopes & 36 & 244 & 1,197 & 4,990 & 17,101 & 50,376 &  73,944\\ [0.3em]
\hline 
 &  &  &  &  &  &  & \\ [-0.8em]
fav. polytopes & 36 & 243 & 1,185 & 4,897 & 16,608 & 48,221 &  71,190 \\ [0.3em]
\hline 
 &  &  &  &  &  &  & \\ [-0.8em]
fav. FRSTs & 48 & 525 & 5,330 & 56,714 & 584,281 & 5,990,333 & 6,637,231 \\ [0.3em]
\hline 
\hline 
 &  &  &  &  &  &  & \\ [-0.8em]
involutions & 184 & 3,035 & 39,653 & 495,854 & 5,777,640 & 65,625,277 & 71,941,643  \\ [0.3em]
\hline 
 &  &  &  &  &  &  & \\ [-0.8em]
smooth invol. & 138 & 1,975 & 22,933 & 230,886 & 2,081,080  & 17,875,122\footnotemark & 20,212,134 \\ [0.3em]
\hline 
\hline 
 &  &  &  &  &  &  & \\ [-0.8em]
only O7 & 49 & 598 & 3,896 & 25,391 & 177,468 & 1,336,960 & 1,544,362  \\ [0.3em]
\hline 
 &  &  &  &  &  &  & \\ [-0.8em]
$\geq 2$ coin. O3 & 71 & 1,089 & 15,497 & 164,634 & 1,480,968 & 12,596,558 & 14,258,817  \\ [0.3em]
\hline 
\end{tabular} 
}
\caption{Number of CY orientifolds obtained in our scan.
We also collected the numbers of the models with only O7-planes or more than one coincident O3-plane.
}\label{tab:ScanData}
\end{table}
\footnotetext{Parts of the orientifold data for $h^{1,1}=7$ are still work in progress and will be updated in the repository as soon as possible.}

The database we produce consists of two sets of data:
\begin{enumerate}
\item We compute \emph{all} FRSTs of \emph{all} favourable polytopes at $h^{1,1}\leq 7$.
For each toric coordinate $z_{k}$,
we construct the orientifold configuration associated with the involution $z_{k}\raw -z_{k}$.
This data is summarised in Tab.~\ref{tab:ScanData}.
\item For each combination of Hodge numbers $(h^{1,1},h^{1,2})$ up to $h^{1,1}=12$,
we generate up to $20$ \emph{random} FRSTs of $\leq 20$ favourable polytopes.
Again,
we build orientifolds for involutions of each toric coordinate $z_{k}\raw -z_{k}$.
The results are given in Tab.~\ref{tab:ScanDataRandom}.
\end{enumerate}
The full data are collected in a GitHub repository which can be found here: \url{https://github.com/AndreasSchachner/CY_Orientifold_database}.

As we said, for each triangulation, we analyse each involution $z_k\mapsto - z_k$ and determine the fixed point set. In Table~\ref{tab:ScanData} and in Table~\ref{tab:ScanDataRandom} we report the numbers of independent\footnote{We count the number of inequivalent involutions given that inverting coordinates with the same weight vector gives rise to equivalent involutions up to coordinate redefinitions.} involutions. Some of these involutions lead to singularities in the CY threefold. As explained at the end of Section~\ref{sec:OrientifoldData}, we can detect the singular models. We refer to models that do not present manifest pathologies as \emph{smooth involutions}.

We finally report the number of models that contain only O7-planes and those that contain at least two O3 planes that can collide by a complex structure deformation of the threefold.
Models in both classes will be suitable for T-brane de Sitter uplift, while models in the last class are needed in order to implement de Sitter uplift by an  anti-D3-brane at the tip of a highly warped throat realising the scenario outlined in \cite{Kallosh:2015nia,Garcia-Etxebarria:2015lif,Crino:2020qwk}.

As observed in \cite{Cicoli:2021dhg},
there is a trend that del Pezzo divisors $\mathrm{dP}_{n}$ with $1\leq n\leq 5$ embedded into CY threefolds obtained from the KS database never satisfy the diagonality condition \eqref{eq:diagdP}, cf.~Tab.~\ref{tab:D3charge}.
Our analysis extends the conjecture of \cite{Cicoli:2021dhg} to all FRSTs at $h^{1,1}=6,7$.

\begin{table}[t!]
\centering
\resizebox{\columnwidth}{!}{
\begin{tabular}{|c|c|c|c|c|c|c||c|}
\hline 
&  &  &  &  &  &  & \\ [-0.8em]
$h^{1,1}$ & 7 & 8 & 9 & 10 & 11 & 12  & total \\ [0.3em]
\hline 
\hline 
 &  &  &  &  &  &  & \\ [-0.8em]
fav. polytopes & 1,219 & 1,498 & 1,587 & 1,555 & 1,623 & 1,807 & 8,980\\ [0.3em]
\hline 
 &  &  &  &  &  &  & \\ [-0.8em]
fav. FRSTs & 4,560 & 6,897 & 9,968 & 12,189 & 15,748 &15,430 & 64,792\\ [0.3em]
\hline 
\hline 
 &  &  &  &  &  &  & \\ [-0.8em]
involutions & 49,326 & 81,911 & 128,403 & 169,775 & 235,216 & 245,989 &  910,620 \\ [0.3em]
\hline 
 &  &  &  &  &  &  & \\ [-0.8em]
smooth invol.   & 6,491 & 9,102 & 13,041 & 15,713 & 21,892 & 24,154 &  90,393 \\ [0.3em]
\hline 
\hline 
 &  &  &  &  &  &  & \\ [-0.8em]
only O7  & 1,769 & 2,608 & 3,493 & 3,543 &4,330 & 4,772 & 20,515 \\ [0.3em]
\hline 
 &  &  &  &  &  &  & \\ [-0.8em]
$\geq 2$ coin. O3  & 3,168 & 4,084 & 5,865 &  6,692 & 9,978 & 9,507& 39,294 \\ [0.3em]
\hline 
\end{tabular} 
}
\caption{Random FRSTs for favourable polytopes between $7\leq h^{1,1}\leq 12$.
We selected up to $20$ polytopes for each combination of Hodge numbers $(h^{1,1},h^{1,2})$ available in the KS database.
}\label{tab:ScanDataRandom}
\end{table}

\subsection{Hodge and Euler numbers of toric divisors}

In this section,
we investigate the divisor data of CY threefolds with $h^{1,1}\leq 6$.
We computed the Hodge numbers of prime toric divisors via the methods described in Sect.~\ref{sec:DivTops} which is largely consistent with the data presented in \cite{Altman:2021pyc}.
We compare the D3-charge contribution of $\mathrm{SO}(8)$ stacks (local D7-tadpole cancellation)  with that of Whitney branes (non-local D7-tadpole cancellation).
We argue that there is an enhancement of about a factor of $5$ between local and non-local D7-tadpole cancellation.

Recalling \eqref{eq:ChiDivisorHodgeNums},
it is clear that divisors with $h^{(0,1)}(D)=0$ lead to the largest Euler characteristic.
That is,
it seems to be profitable to have O7-planes and D7-branes wrapping divisors with Hodge numbers
\begin{equation}
h^{\bullet}(D)=\lbrace 1, 0, h^{(0,2)}(D), h^{(1,1)}(D)\rbrace\:,
\end{equation}
for which \eqref{eq:ChiDivisorHodgeNums} leads to
\begin{equation}
\chi(D)=2\left (1+h^{2,0}(D)\right )+h^{1,1}(D)\, .
\end{equation}

\begin{figure}[t!]
\centering
\includegraphics[scale=0.5]{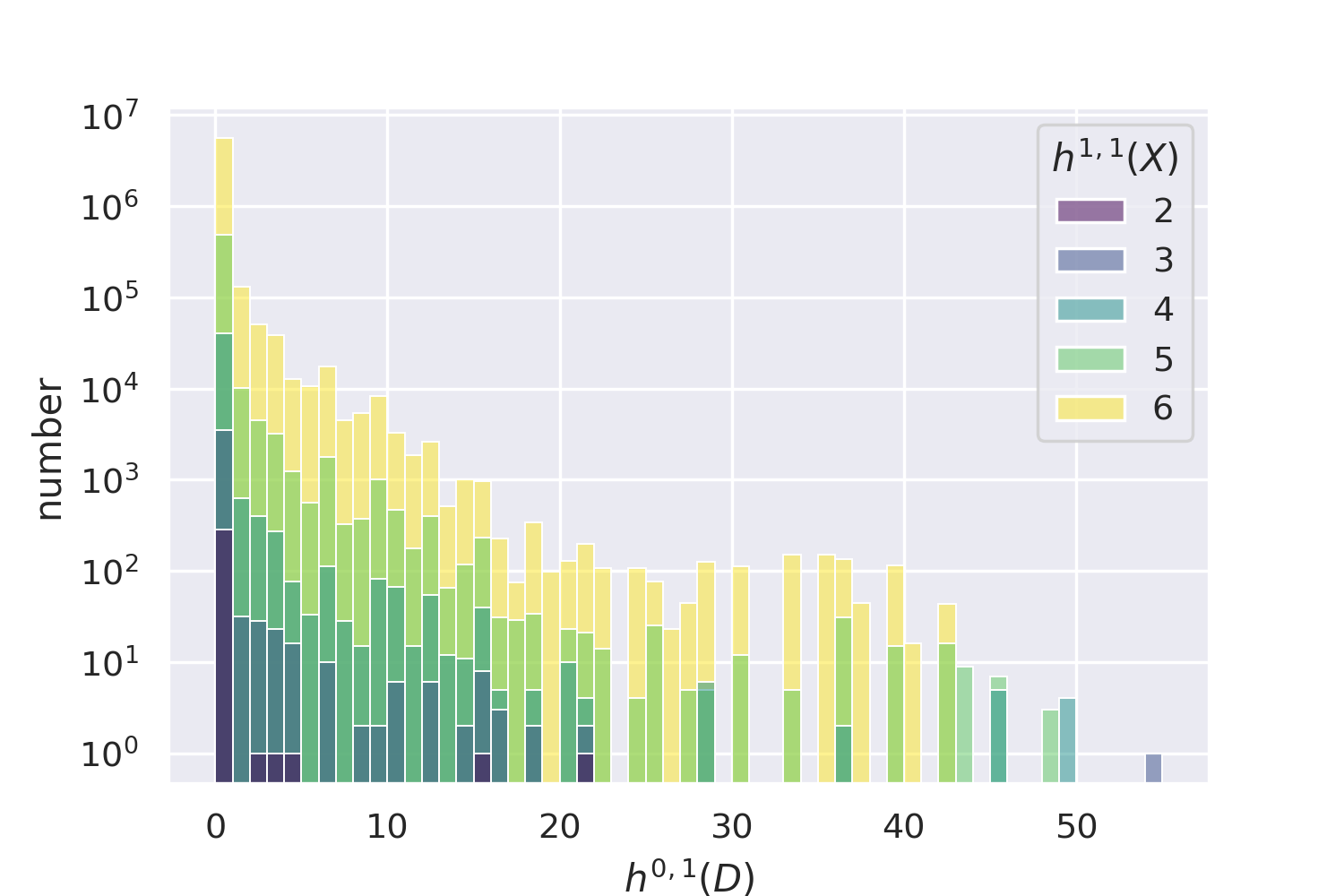}\includegraphics[scale=0.5]{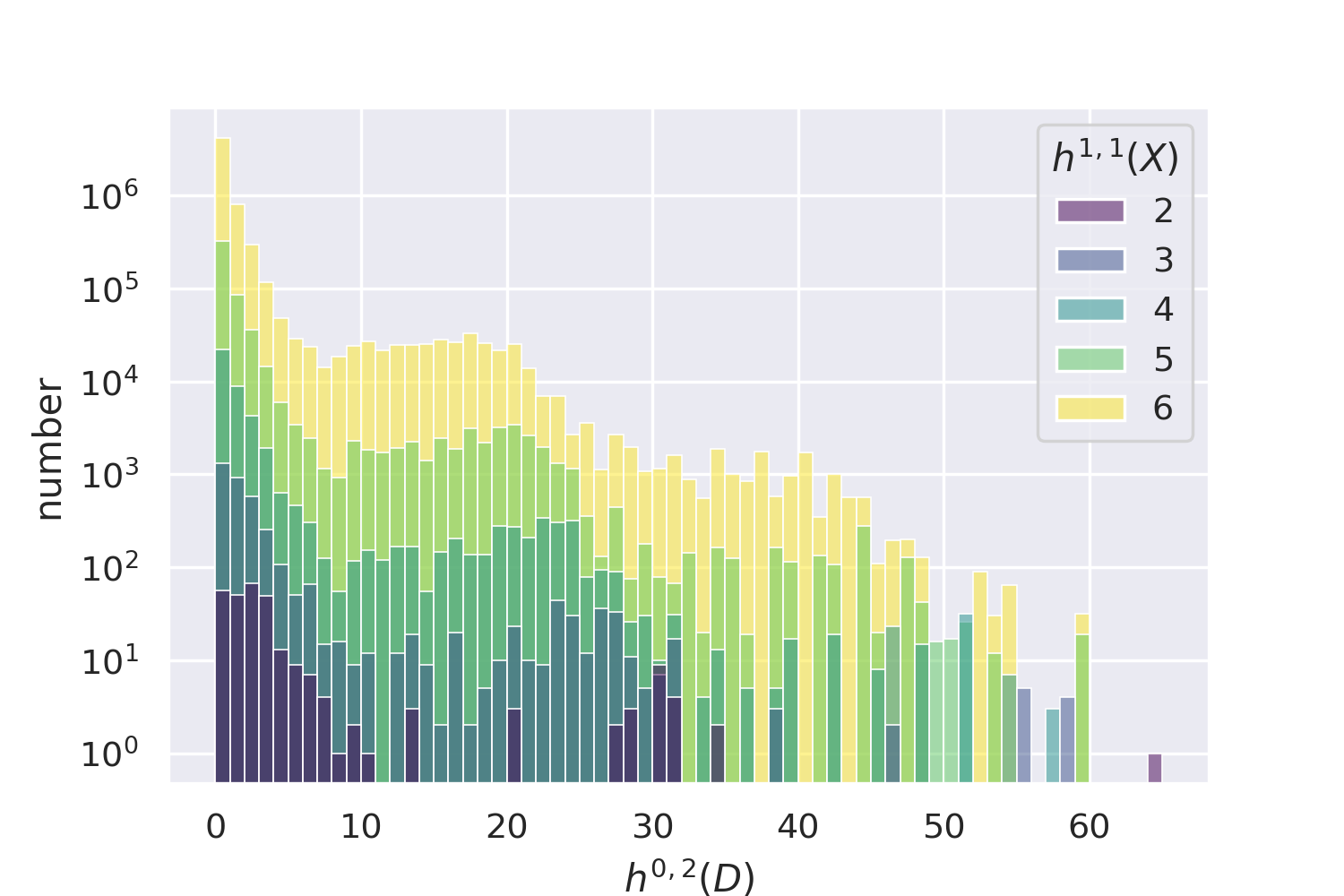}
\includegraphics[scale=0.5]{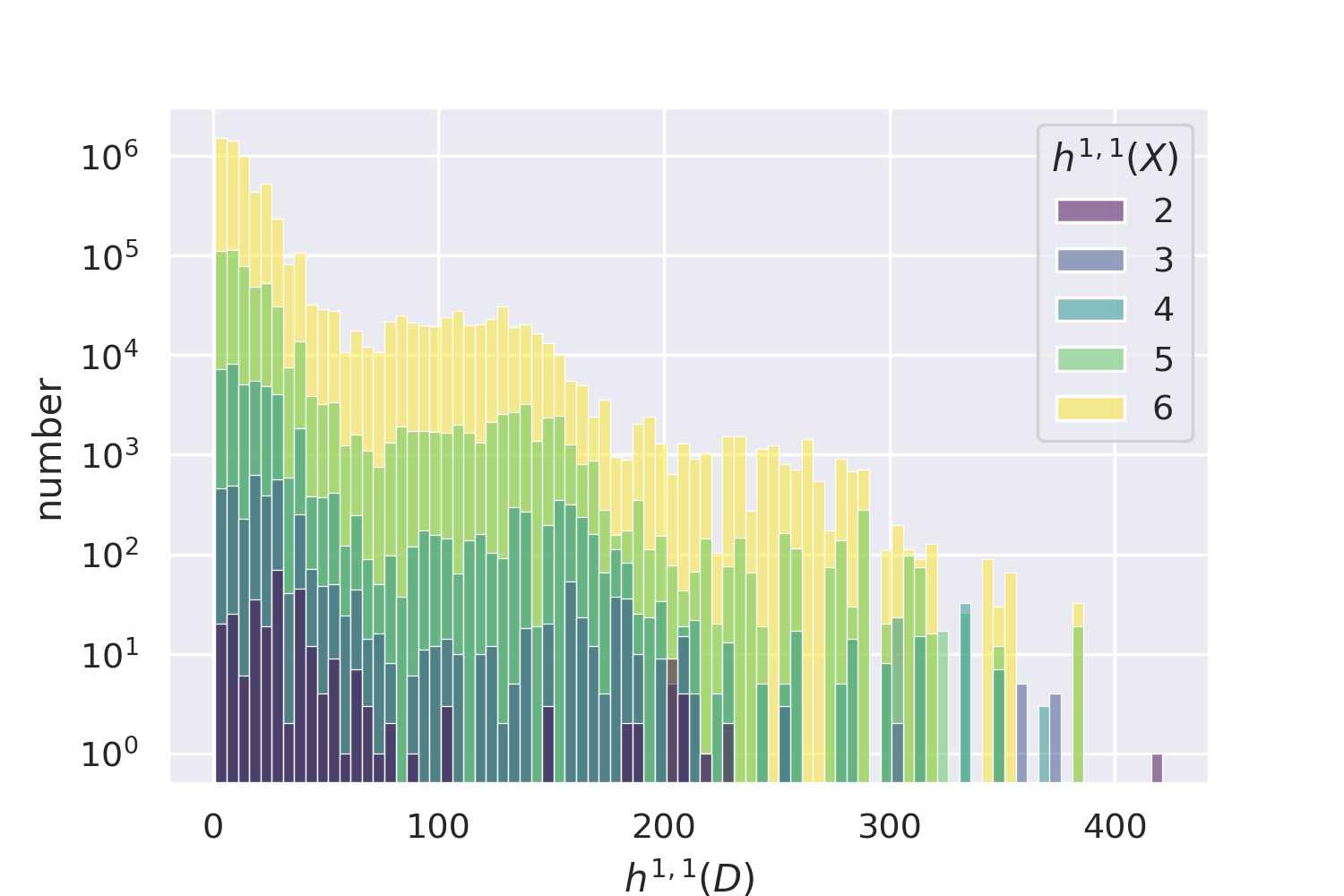}\includegraphics[scale=0.55]{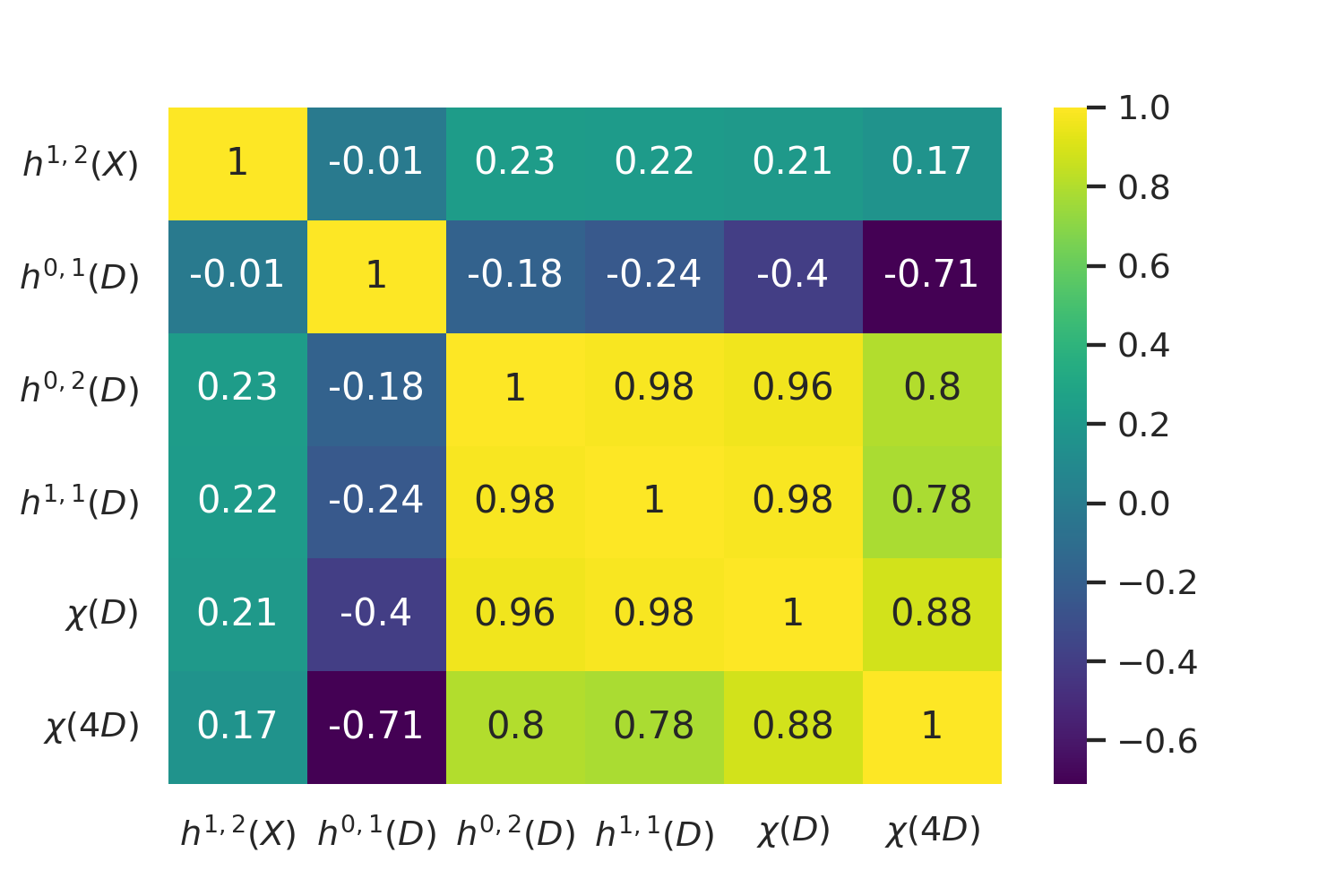}
\caption{Histogram plots for Hodge numbers $h^{p,q}$ of all divisors at $h^{1,1}\leq 6$.
We ignore $h^{0,0}$ given that $h^{0,0}(D)=1$ for all $D$.
The bottom right plot shows a correlation map for the relevant divisor and CY data.}\label{fig:Hist1} 
\end{figure}

\begin{figure}[t!]
\centering
\includegraphics[scale=0.5]{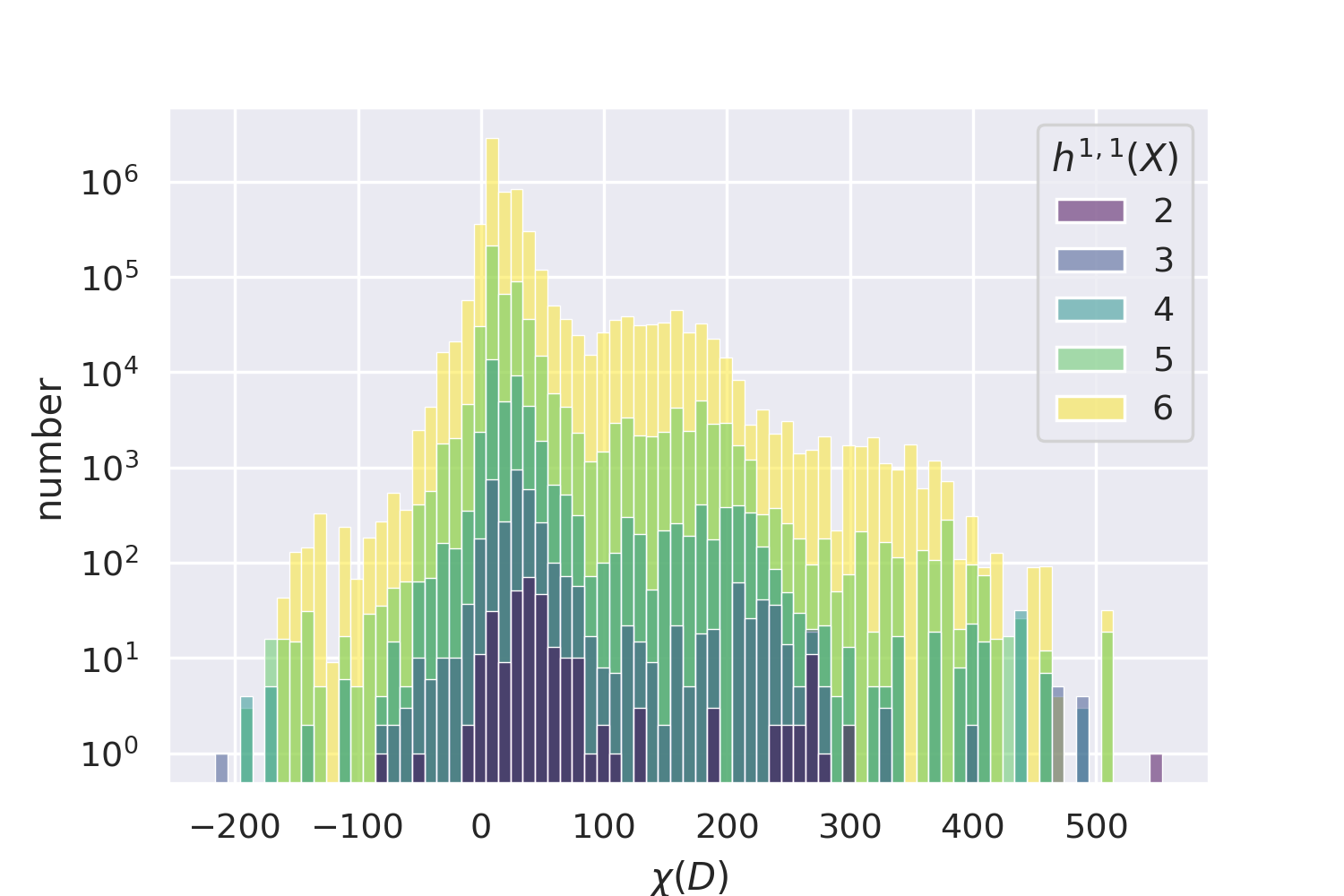}\includegraphics[scale=0.5]{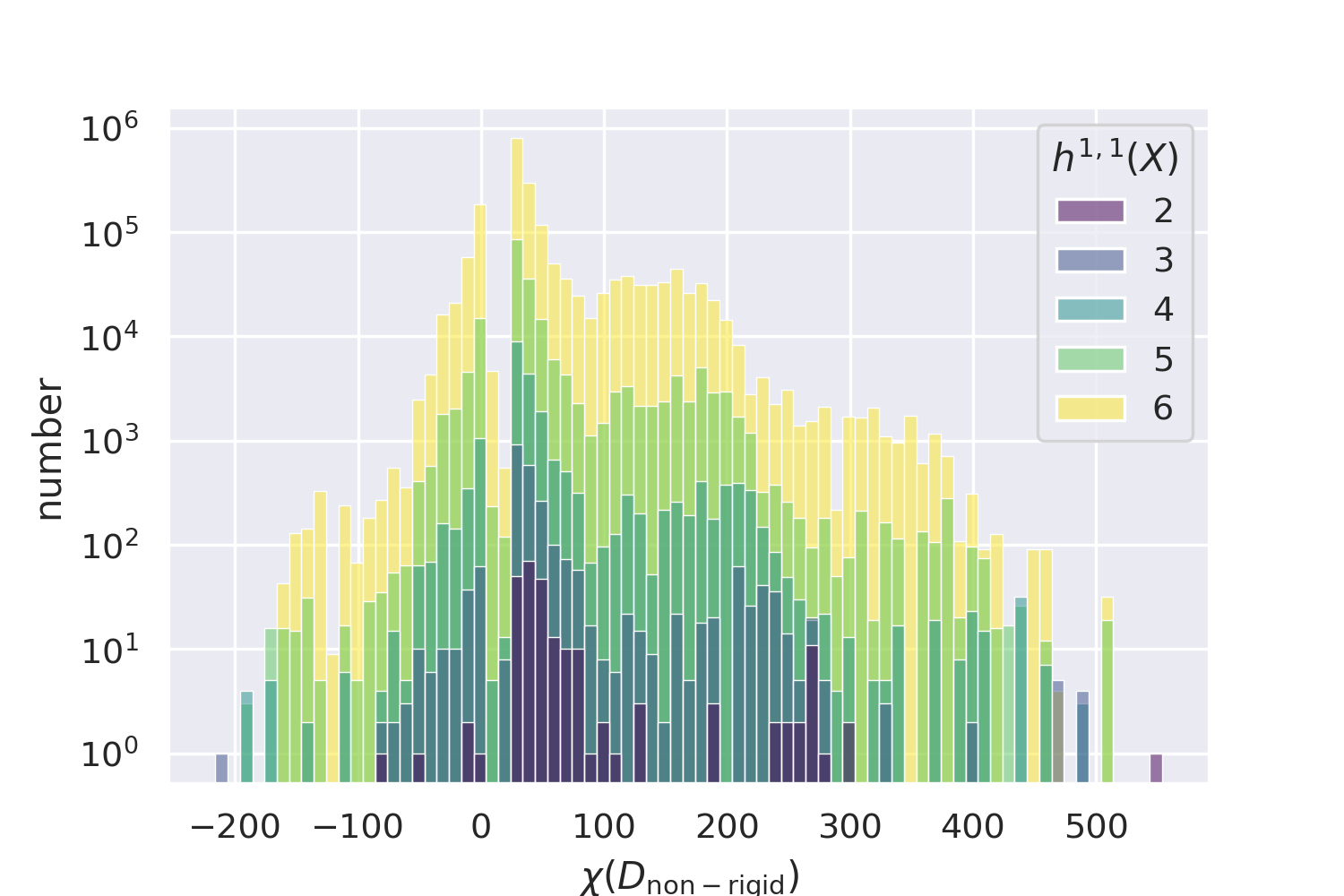}
\includegraphics[scale=0.5]{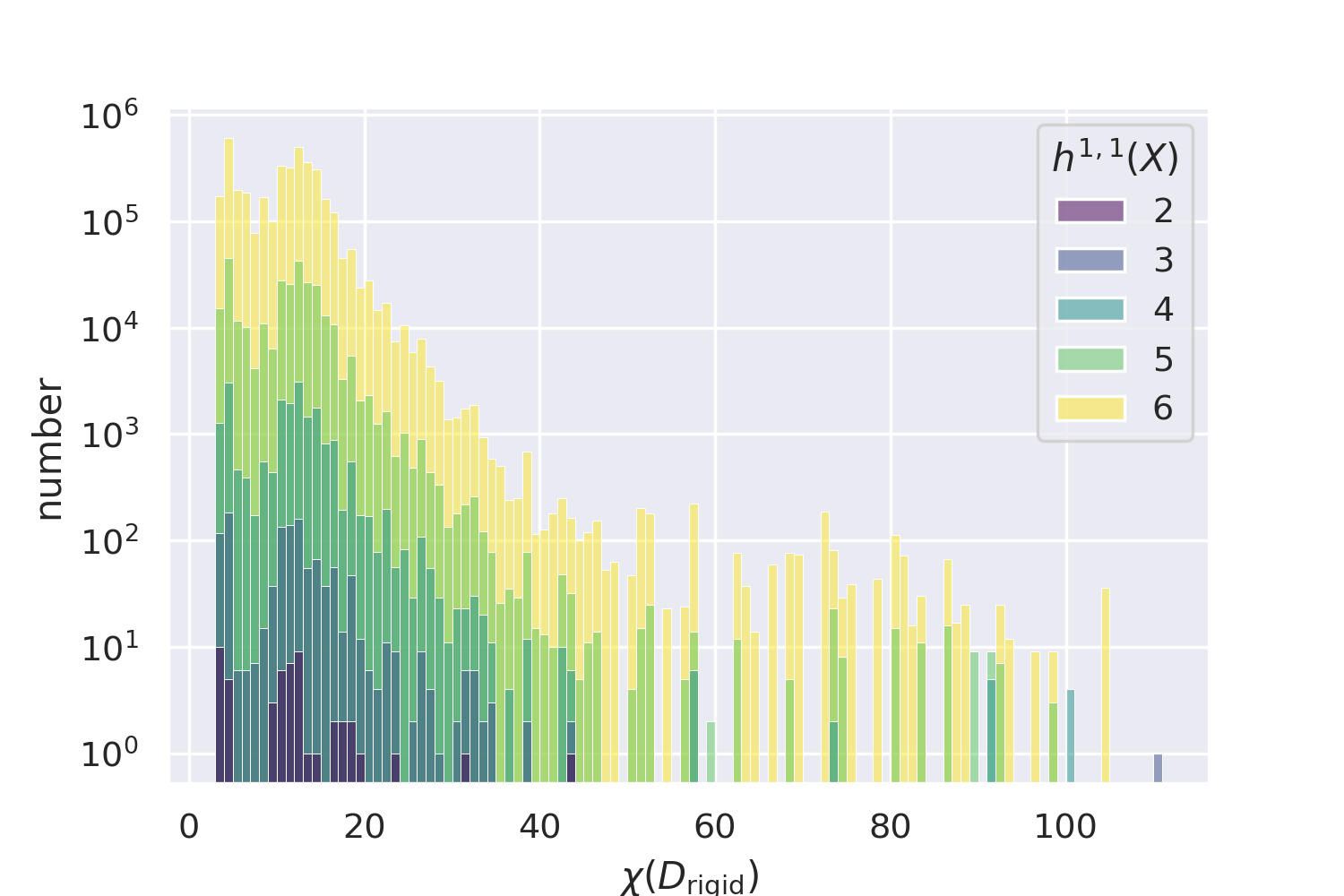}\includegraphics[scale=0.5]{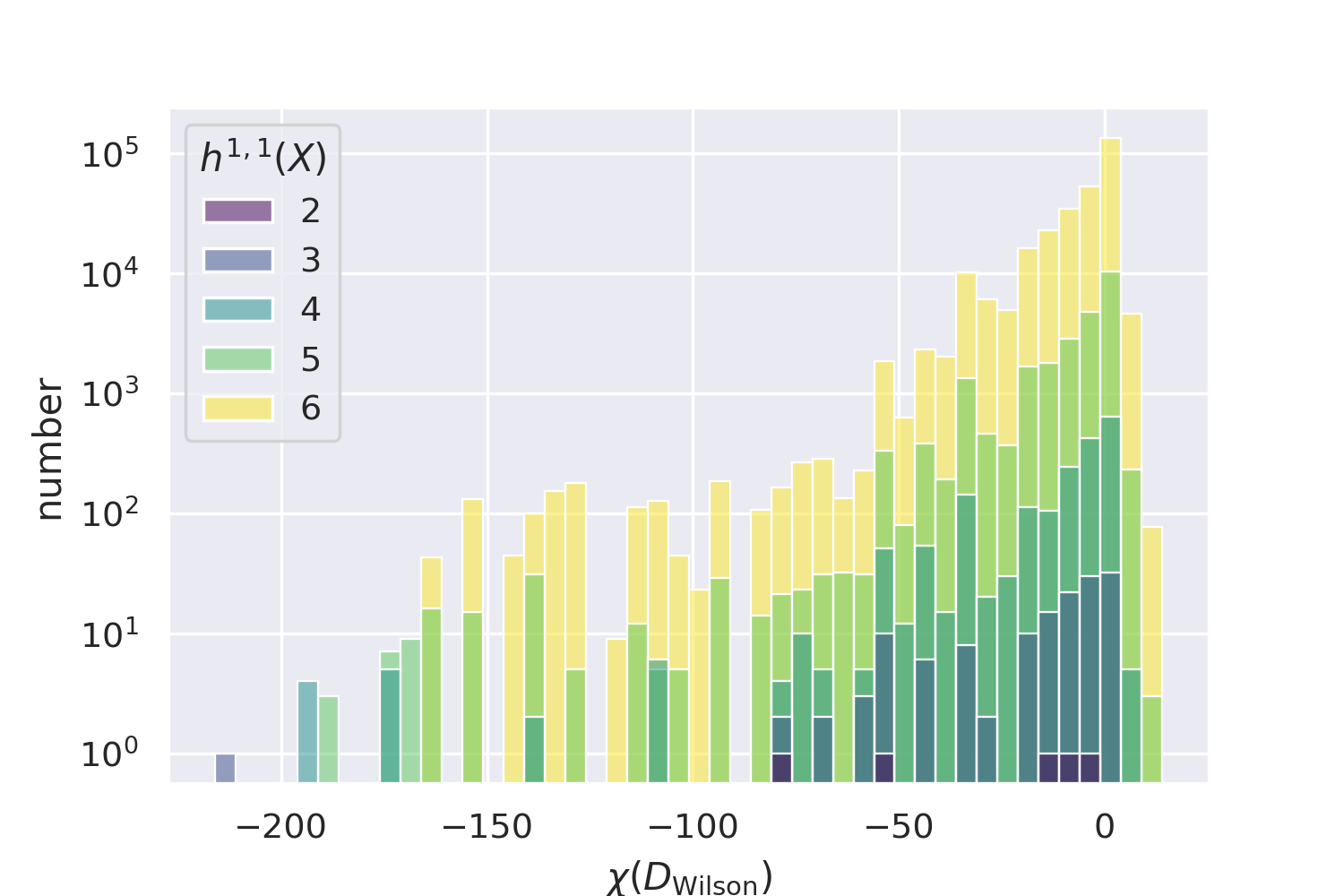}
\caption{Histogram plots for Euler numbers of divisors.
In the first row, we computed the distribution of Euler numbers for all divisors $D$ on the left and for only non-rigid divisors $D_{\text{nr}}$ with either $h^{0,1}(D)>0$ or $h^{0,2}(D)>0$.
The second row shows the Euler characteristic for divisors which are rigid (left) or Wilson (right) with $h^{1,0}\neq 0$, $h^{2,0}=0$.}\label{fig:Hist2} 
\end{figure}

Clearly,
maximising $\chi(D)$ is beneficial from the perspective of the tadpole \eqref{eq:DThreeTadpoleGen}.
Using the results computed in our database for $h^{1,1}\leq 6$,
we compute the Euler characteristic for every divisor finding that the maximal value is\footnote{For models that admit exchange involutions, one verifies that $\chi(D)\bigl |_{\text{max}}=492$ in agreement with \cite{Gao:2022fdi}.}
\begin{equation}\label{eq:MaxChiH11_6} 
\chi(D)\bigl |_{\text{max}}=549\, .
\end{equation}
A complete overview of the distribution of both Hodge numbers as well as Euler characteristics of (prime toric) divisors appearing in the KS database up to $h^{1,1}=6$ is shown in Fig.~\ref{fig:Hist1} and Fig.~\ref{fig:Hist2} respectively.

In Fig.~\ref{fig:Hist2},
we show the distribution of Euler numbers for different types of divisors.
Clearly, non-rigid divisors result in the largest $\chi(D)$ with the maximum given by \eqref{eq:MaxChiH11_6}, while $\chi(D_{\text{rigid}})\bigl |_{\text{max}}=111$ for rigid divisors.
Those divisors $D$ with non-positive $\chi(D)$ are in fact associated with Wilson divisors\footnote{
We compared these results to the Hodge numbers obtained from the database \cite{Altman:2021pyc} and found overall agreement.} with $h^{0,1}(D)>0$ and $h^{0,2}(D)=0$ out of which $44.67\%$ are Exact-Wilson divisors with $h^{0,1}(D)=1$.

\

In Fig.~\ref{fig:Hist1},
we present a correlation map for Hodge and Euler numbers of divisors.
The correlations between $\chi(D)$ and the corresponding Hodge numbers is clear from \eqref{eq:ChiDivisorHodgeNums}.
In our data,
there are no significant correlations between any of the variables shown on the bottom right of Fig.~\ref{fig:Hist1} with the number of Kähler moduli $h^{1,1}(X)$ of the CY $X$ nor with $h^{0,0}(D)$ which is why we omitted the later two.
Interestingly,
we observe that there is an anti-correlation between $h^{0,1}(D)$ with $h^{0,2}(D)$ and $h^{1,1}(D)$, while at the same time $h^{0,2}(D)$ and $h^{1,1}(D)$ are strongly correlated.
This implies that there is an obvious trend where larger $h^{1,1}(D)$ implies large $h^{0,2}(D)$ plus small $h^{0,1}(D)$ and, hence, larger $\chi(D)$.
We observe similar correlations of $\chi(D)$ and $\chi(4D)$ with $h^{1,2}(X)$ which will be confirmed further below in Fig.~\ref{fig:OverviewD3ChargeOp}.

\

\newpage

\begin{figure}[t!]
\centering
\includegraphics[scale=0.7]{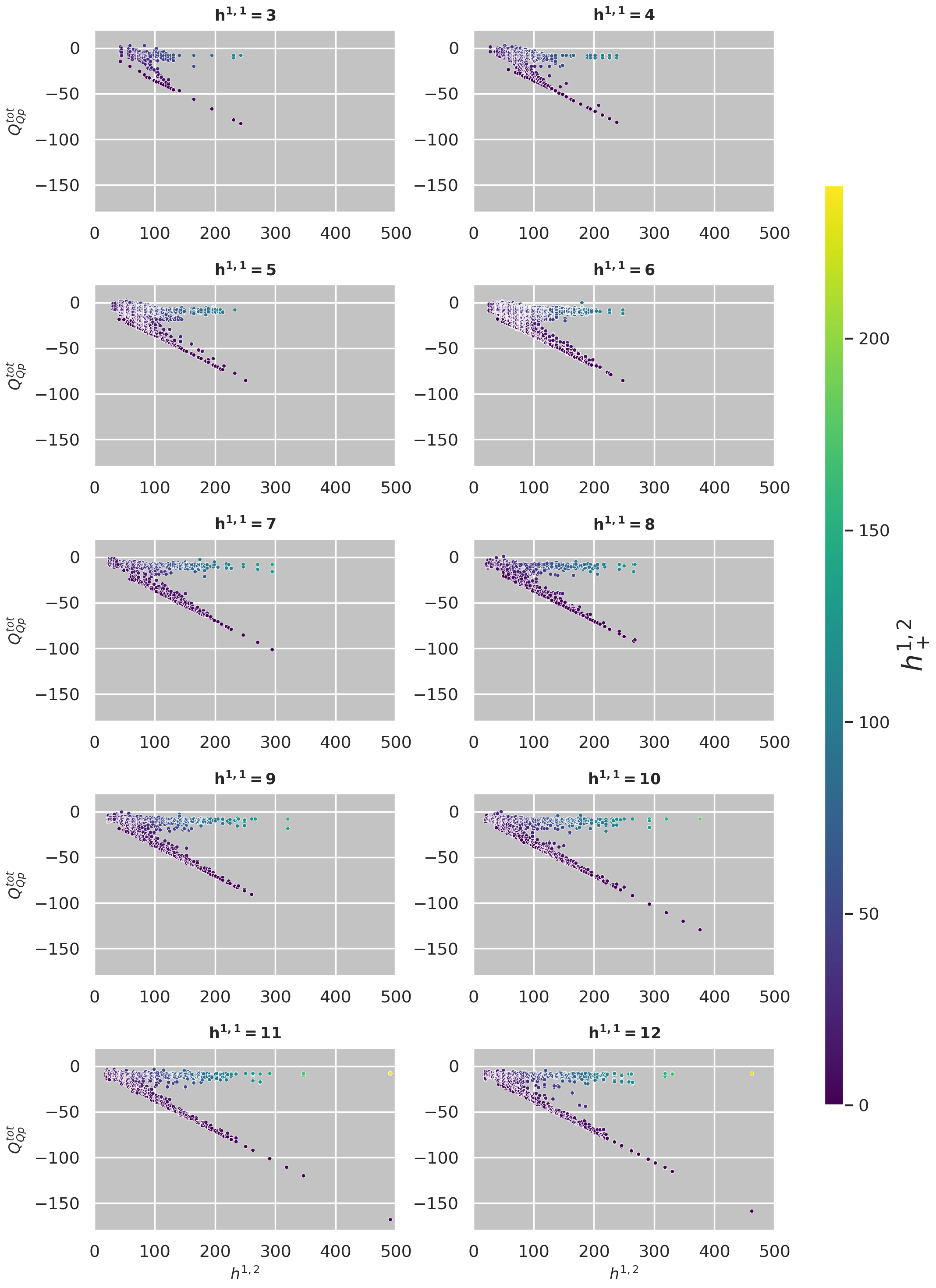}
\caption{Overview of the D3-charge contribution contributed by O$p$-planes only.
For $h^{1,1}\geq 7$,
we present the data for models collected in Tab.~\ref{tab:ScanDataRandom}, while for $h^{1,1}\leq 6$ we use the models of Tab.~\ref{tab:ScanData}.
The maximal absolute value for the D3-charge comes from orientifolds with Hodge numbers $(h^{1,1},h^{1,2}_{-},h^{1,2}_{+})=(11,491,0)$ where we find $Q_{Op}^{\text{tot}}=-168$.}\label{fig:OverviewD3ChargeOp} 
\end{figure}

\begin{landscape}
\begin{table}[t!]
\centering
\resizebox{1\columnwidth}{!}{
\begin{tabular}{|c||c|c|c|c|c|c|c|c|c|c|c|c|}
\hline 
&  &  &  &  &  &  &  &  &  &  &  &  \\ [-0.8em]
$h^{1,1}$&   & 2 & 3 & 4 & 5 & 6 & 7& 8& 9& 10& 11& 12   \\ [0.3em]
\hline 
\hline 
\multirow{5}{*}[1.em]{D3-charges $|Q_{\text{D3}}|$} &&  &  &  &  &  &  &  &  &  &  &  \\ [-0.8em]
& local& 276 & 248 & 244 & 256 & 256 & 304 & 276 & 272 & 388 & 504 &  476   \\ [0.3em]
\cline{2-13}
 &&  &  &  &  &  &   &  &  &  &  &  \\ [-0.8em]
& non-local& 3,678 &  3,272 & 3,212 &  3,280 & 3,280 & 4,000 & 3,594 & 3,408 & 5,036 & 6,664 & 6,258  \\ [0.3em]
\hline 
\hline 
\multirow{18}{*}[1em]{dP$_n$ (ddP$_n$)}
 &&  &  &  &  &  &  &  &  &  &  &  \\ [-0.8em]
&dP$_0$ (ddP$_0$) & 10 (*) & 117 (*) & 1,282 (*) & 15,346 (*) & 172,469 (*) & 526 (*)  & 656 (*) & 1,135 (*) & 1,049 (*) & 1,424 (*) & 2,086 (*)  \\ [0.3em]
\cline{2-13}
 &&  &  &  &  &  &   &  &  &  &  &  \\ [-0.8em]
& dP$_1$ (ddP$_1$)  & 5 (3) & 159 (47) & 2,677 (726) & 39,355 (8,880) & 514,099 (93,000) & 4,994 (59) & 8,743 (34) & 15,007 (61) & 20,346 (35) & 28,962 (87) & 31,671 (49)   \\ [0.3em]
\cline{2-13}
 &&  &  &  &  &  &   &  &  &  &  &  \\ [-0.8em]
&dP$_2$ (ddP$_2$) & 0 (0) & 6 (0) & 438 (0) & 10,926 (0) & 184,992 (0) & 2,822 (0)  & 6,060 (0) & 10,176 (0) & 15,104(0) & 22,350 (0) & 23,178 (0)  \\ [0.3em]
\cline{2-13}
 &&  &  &  &  &  &  &  &  &  &  &   \\ [-0.8em]
&dP$_3$ (ddP$_3$)& 0 (0) & 6 (0) & 359 (0) & 9,211 (0) & 166,494 (0) & 2,182 (0) & 4,375 (0) & 7,004 (0) & 10,850 (0) & 15,618 (0) & 17,495 (0)   \\ [0.3em]
\cline{2-13}
 &&  &  &  &  &  &  &  &  &  &  &   \\ [-0.8em]
&dP$_4$ (ddP$_4$)& 0 (0) & 2 (0) & 78 (0) & 2,227 (0) & 50,821 (0) & 1,119 (0) & 2,196 (0) & 3,894 (0) &  5,851 (0) & 8,533 (0) & 8,613 (0)   \\ [0.3em]
\cline{2-13}
 &&  &  &  &  &  &  &  &  &  &  &   \\ [-0.8em]
&dP$_5$ (ddP$_5$)& 0 (0) & 15 (0) & 524 (0) & 9,482 (0) & 144,966 (0) & 1,499 (0) & 2,487 (0) & 3,808 (0) & 5,612 (0) & 7,689 (0) & 8,226 (0)   \\ [0.3em]
\cline{2-13}
 &&  &  &  &  &  &  &  &  &  &  &   \\ [-0.8em]
&dP$_6$ (ddP$_6$)& 3 (3) & 37 (31) & 418 (201) & 5,714 (1,214) & 81,636 (4,719) & 959 (0) & 1,611 (0) & 2,492 (0) &3,412 (0)  & 5,569 (20) & 5,839 (0)   \\ [0.3em]
\cline{2-13}
 &&  &  &  &  &  &  &  &  &  &  &   \\ [-0.8em]
&dP$_7$ (ddP$_7$)& 6 (6) & 134 (92) & 2,060 (939) & 26,032 (6,158) & 302,879 (38,692) & 1,841 (23) & 2,686 (12) & 4,063 (21) & 4,822 (9) & 6,735 (0) & 6,933 (12)   \\ [0.3em]
\cline{2-13}
 &&  &  &  &  &  &  &  &  &  &  &   \\ [-0.8em]
&dP$_8$ (ddP$_8$)& 7 (7) & 134 (83) & 1,806 (584) & 22,442 (3,458) & 269,626 (24,109) & 1,722 (41) & 2,356 (82) & 3,283 (91) & 4,149 (67) & 4,872 (21) & 4,654 (66)   \\ [0.3em]
\hline 
\hline 
\multirow{8}{*}[1em]{divisor topologies}
 &&  &  &  &  &  &  &  &  &  &  &   \\ [-0.8em]
& {\small rigid} & 52 & 1,164 & 20,339 & 297,112 & 3,840,467 & 34,423 & 58,138 & 94,621 & 126,190 & 178,320 & 187,751   \\ [0.3em]
\cline{2-13}
&&  &  &  &  &  &   &  &  &  &  & \\ [-0.8em]
& {\small $h^{0,2}=1$} & 50 & 924 & 8,931 & 86,418 & 798,972 & 6,065  & 8,431 & 10,555 & 12,941 & 15,501 & 15,550  \\ [0.3em]
\cline{2-13}
 &&  &  &  &  &  &   &  &  &  &  &  \\ [-0.8em]
& {\small $h^{0,2}>1$} & 181 & 1,444 & 11,486 & 101,911 & 910,903 & 7,689 & 12,121 & 18,008 & 22,478 & 29,611 &  27,825   \\ [0.3em]
\cline{2-13}
 &&  &  &  &  &  &   &  &  &  &  &  \\ [-0.8em]
& {\small Wilson } & 5 & 143 & 1,884 & 24,985 & 292,468 & 1,983  & 4,074 & 6,400 & 9,037 & 12,788 & 15,754  \\ [0.3em]
\hline 
\end{tabular} 
}
\caption{Top: Minimal D3-charge contributions from O$p$-planes and the corresponding total for either local or non-local D7-tadpole cancellation.
Middle: Distribution of $\mathrm{dP}_{n}$ and diagonal $\mathrm{ddP}_{n}$ divisors.
dP$_0=\mathbb{P}^{2}$ always appears diagonally for which we write $(*)$.
For $h^{1,1}=7$, we present the data obtained from random runs.
Bottom: Summary of topologies of toric divisors encountered in the scan.
In the second and third row, we distinguish between deformation divisors ($h^{0,1}=0$) with $h^{0,2}=1$ and $h^{0,2}>1$.
}\label{tab:D3charge} 
\end{table}
\end{landscape}

\subsection{D3-charge in the database and non-local D7-tadpole cancellation}

In this section, we will study how the D3-tadpole contribution from localised sources varies in the dataset we are taking into account.

We begin by considering only the D3-charge coming from the O-planes.
We present an overview of their D3-charge contribution in Fig.~\ref{fig:OverviewD3ChargeOp}.
We ignored models with positive $Q_{\text{O}p}^{\text{tot}}$.
The colouring indicates the value of $h^{1,2}_{+}$ where we clearly see the trend expected from \eqref{eq:D3OpHodgeNum}:
non-vanishing $h^{1,2}_{+}$ decreases the absolute value of the D3-charge contribution.
The models on the diagonal line have $h^{1,2}_{+}=0$ and follow the expected scaling $\sim -h^{1,2}_{-}/3$ as derived in \eqref{eq:D3OpHodgeNum}.

Let us introduce the D7-branes. We analyse the situation in the absence of gauge flux on the D7-branes.\footnote{Freed-Witten anomaly cancellation may force some flux to be non-zero; however one can always choose a flux that minimise its contribution to the D3-charge; in this situation our results are good approximations for the total D3-charge coming from localised sources.} 
For each model (derived from a choice of CY $X$ and involution), we try to cancel the D7-tadpole generated by the O7-planes by a D7-brane configuration that maximizes their (absolute value of the) contribution to the D3-charge.
For each O7-plane that we find we then work out the topology of the wrapped divisor. 
If we have O7-planes on rigid or Wilson divisors, we cancel the D7-tadpole by a $\mathrm{SO}(8)$ stacks.
For O7-planes on deformation divisors with $h^{0,2}(D)>1$,
we cancel the D7-tadpole non-locally through Whitney branes, see App.~\ref{app:EWB} for details. Finally, 
whenever $h^{0,2}(D)=1$,
we construct \eqref{eq:DefEqWhitneyBrane} explicitly to check for eventual factorisation; if no factorization is forced, we add a Whitney brane.

For each $h^{1,1}(X)$, we pick the model ($X$ and involution) whose localised sources contribute most to the total D3-charge. 
In Tab.~\ref{tab:D3charge}, we report the absolute value of the total D3-charge from these localised sources for two cases: 1) the D7-tadpole is canceled by putting $4+4$ D7-branes on top of all the O7-planes (local D7-tadpole cancellation) and 2) we put Whitney branes on all non-rigid O7-plane divisors (non-local D7-tadpole cancellation).

\begin{figure}[t!]
\centering

\textbf{Complete scan at $h^{1,1}(X)\leq 6$}

\vspace{-0.1cm}

\includegraphics[scale=0.525]{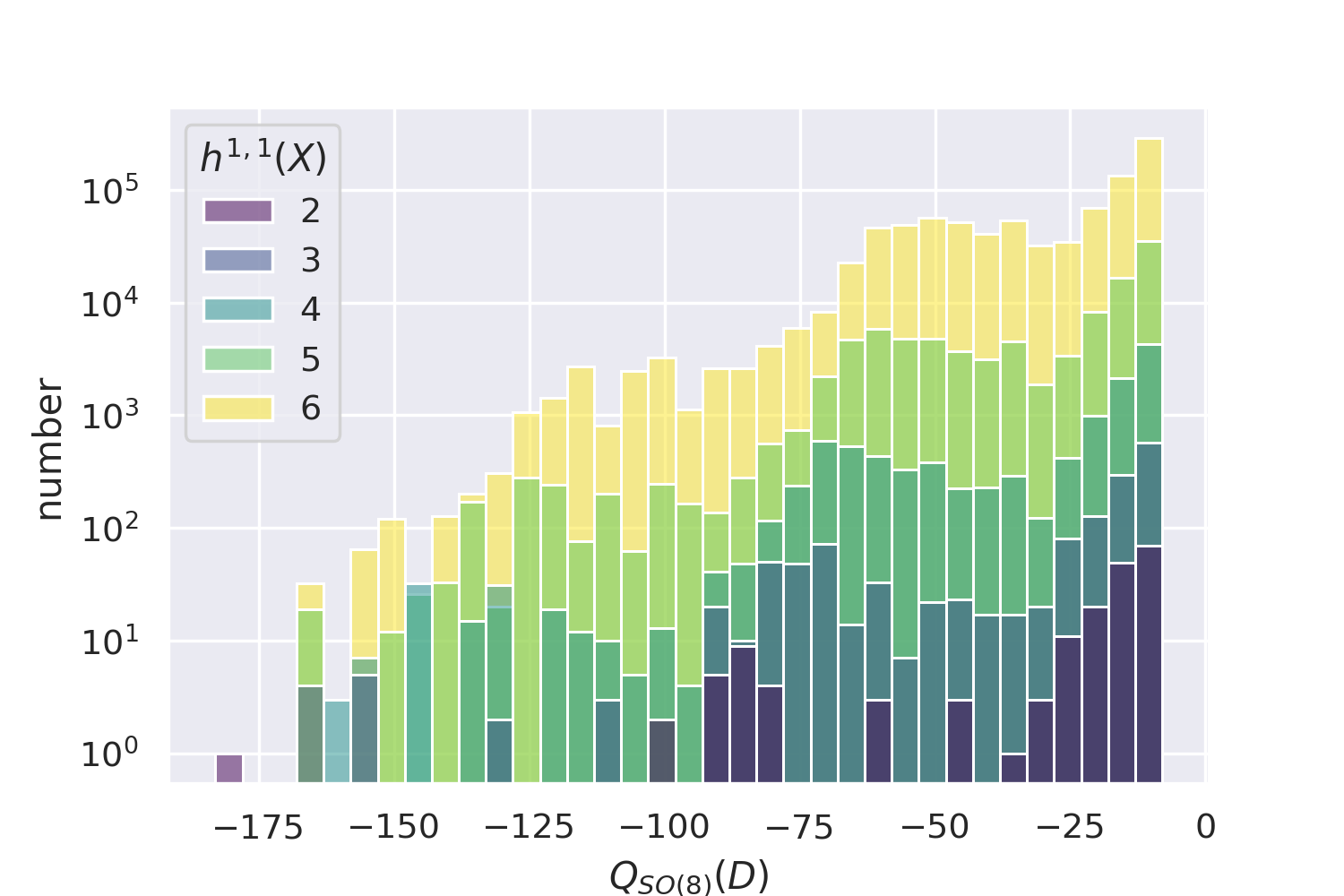}\includegraphics[scale=0.525]{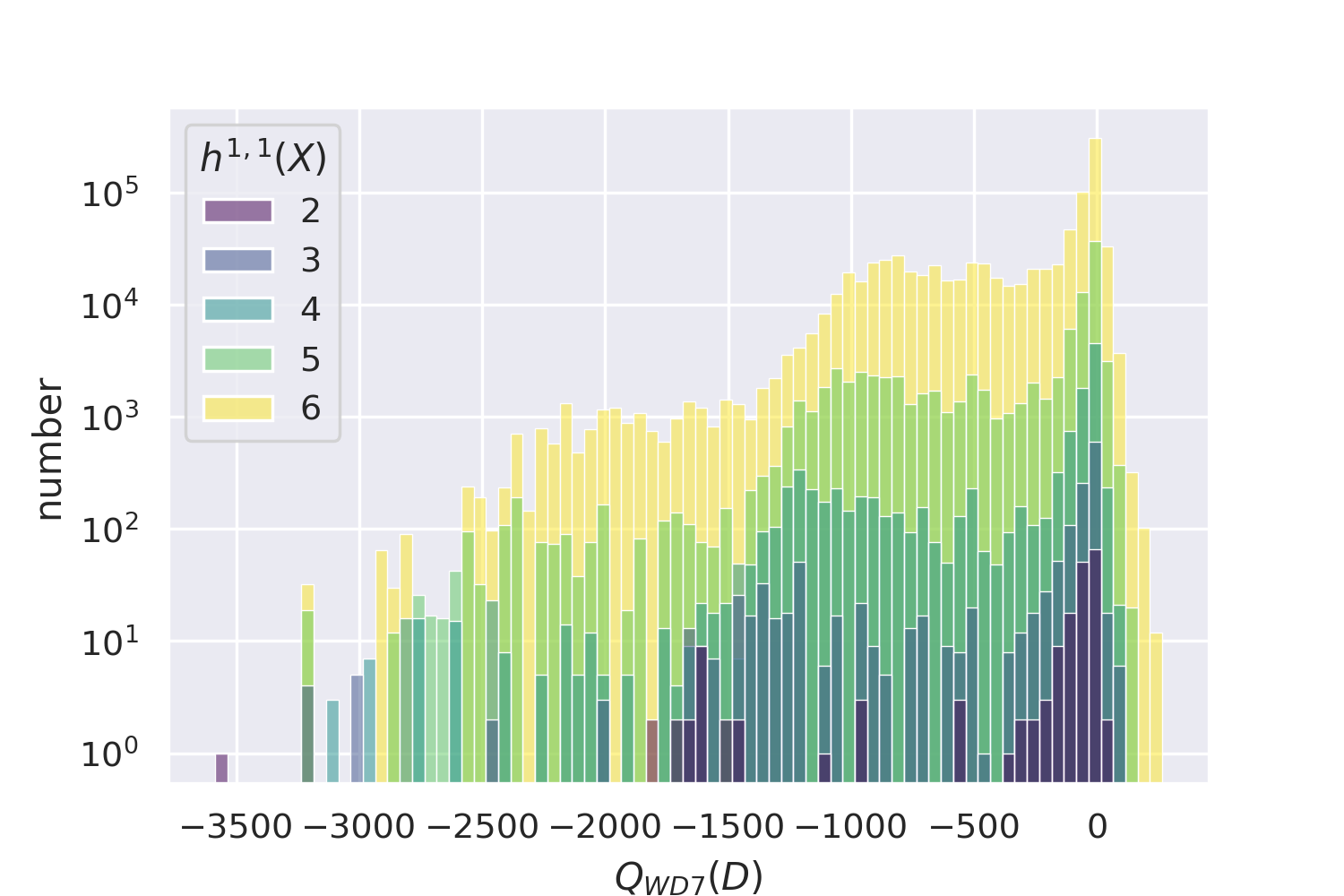}

\vspace{0.2cm}

\textbf{Random data at $7\leq h^{1,1}(X)\leq 12$}

\vspace{-0.1cm}

\includegraphics[scale=0.525]{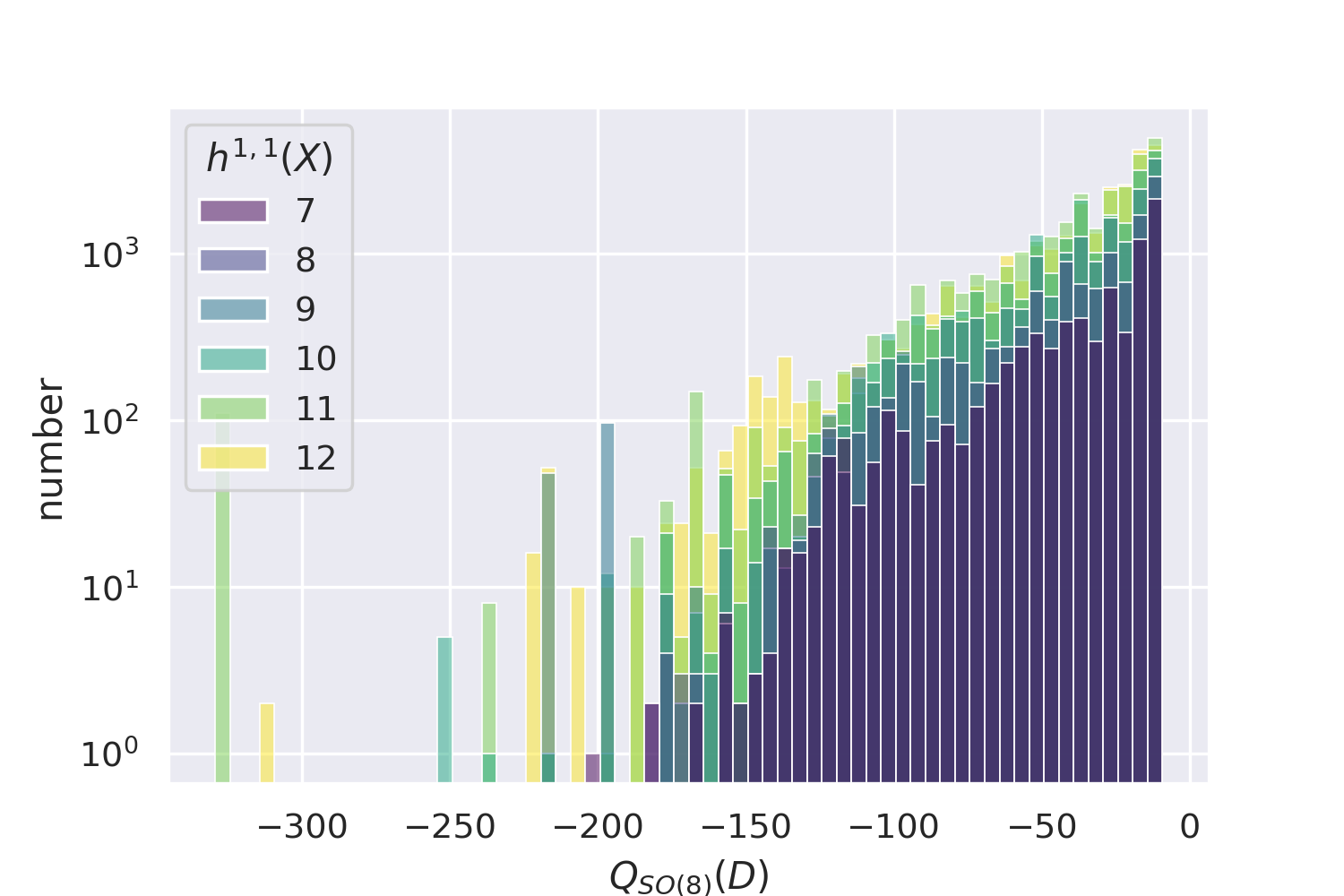}\includegraphics[scale=0.525]{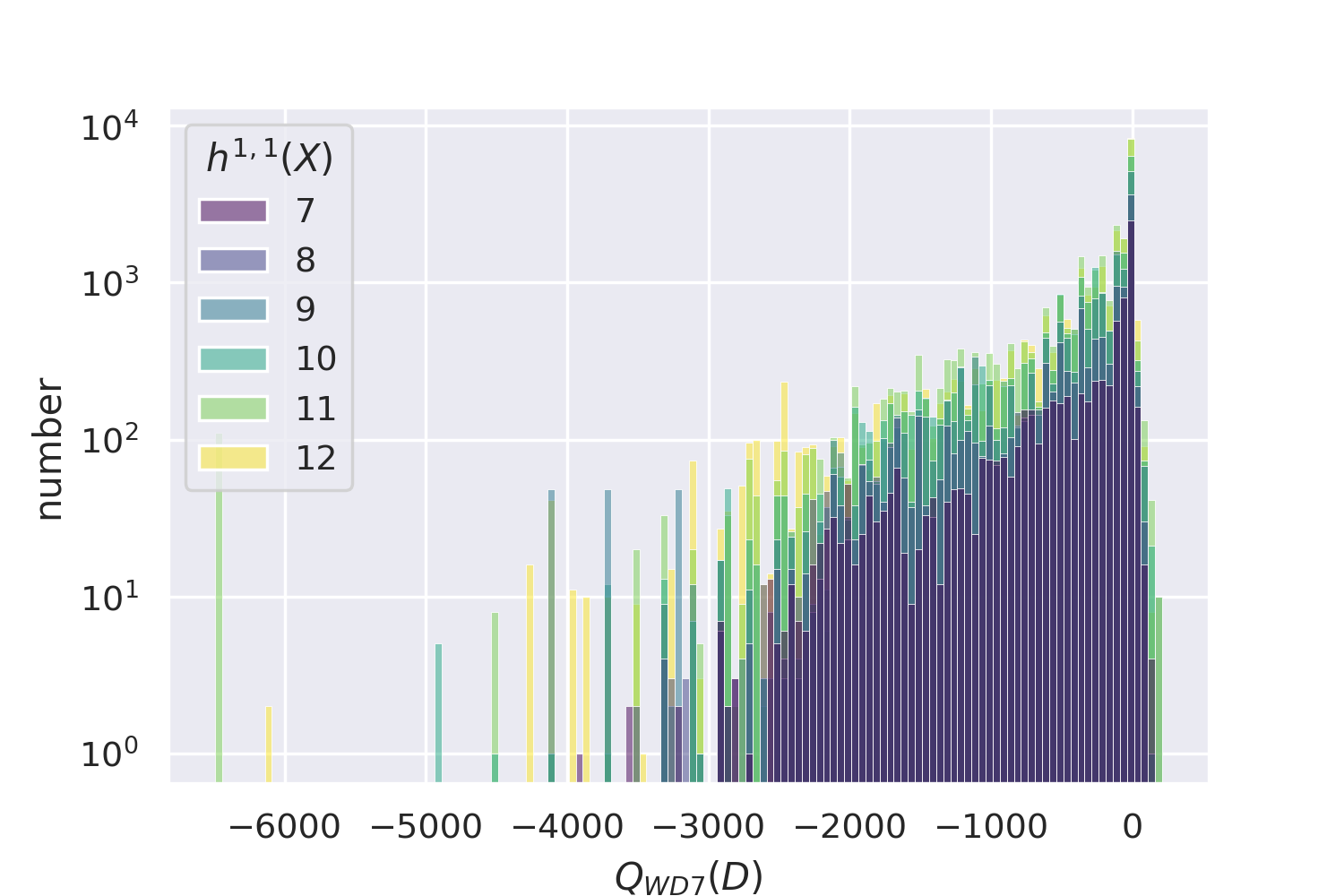}

\caption{Local vs. non-local tadpole cancellation for data at $h^{1,1}(X)\leq 6$. Left: Distribution of $Q_{\text{SO}(8)}(D)$ for divisors with $h^{0,2}(D)>1$ and $h^{0,1}(D)=0$.
Right: Distribution of $Q_{\text{WD7}}(D)$ for the same divisors.}\label{fig:ChiWD7} 
\end{figure}

Let us stress the difference between local and non-local D7-tadpole cancellation.
If we were to simply add (4+4) D7-branes on top of each of the D7-branes to cancel the D7-tadpole locally,
this would amount to\footnote{We ignore the contribution from O3-planes here. For models with the minimal $Q_{\text{O}p}^{\text{tot}}$ on the diagonal in Fig.~\ref{fig:OverviewD3ChargeOp}, there are actually no O3-planes which justifies the bound given in \eqref{eq:UpperBoundLocalTadpoleCanc}.}
\begin{equation}
Q_{\text{D3}}\approx \underbrace{-(4+4)\dfrac{\chi(D_{i})}{24}}_{\text{D7}}\underbrace{-\dfrac{\chi(D_{i})}{6}}_{\text{O7}}\:,
\end{equation}
which leads to the conservative estimate
\begin{equation}\label{eq:UpperBoundLocalTadpoleCanc} 
\text{local D7-tadpole cancellation:  }|Q_{\text{D3}}|\leq 504\:,
\end{equation}
as one can check in Tab.~\ref{tab:D3charge}.
This is precisely the upper bound obtained in \eqref{eq:BoundD3ChargeLocD7} for $(h^{1,1},h^{1,2})=(11,491)$.
In Fig.~\ref{fig:ChiWD7}, we show that the D3-tadpole is significantly enhanced by considering more generic brane configurations, as we argued in Sect.~\ref{sec:D3Tadpole}.
The results in Tab.~\ref{tab:D3charge} show that in this case the total D3-charge extraordinarily exceeds the bound \eqref{eq:UpperBoundLocalTadpoleCanc}.
In particular, as we argue below in Sect.~\ref{sec:H11_11Example},
using instead Whitney branes, the total D3-charge is increased by about a factor of $13$, obtaining the following bound on localised sources D3-charge:
\begin{equation}\label{eq:UpperBoundNonLocalTadpoleCanc} 
\text{non-local D7-tadpole cancellation:  }|Q_{\text{D3}}|\leq 6,664\:.
\end{equation}
The values stated in Tab.~\ref{tab:D3charge} give an upper bound on the total D3-charge.
For models with multiple O3-planes at the tip of a throat which are suitable for anti-D3 uplift \cite{Garcia-Etxebarria:2015lif},
we obtain $|Q_{\text{D3}}|\leq 3592$ with the maximal value realised for orientifolds with Hodge numbers $(h^{1,1}_{+}, h^{1,2}_{-})=(12,274)$.

We finally note that we neglected the D3-contribution from fluxes. The fluxes change the total $Q_{\text{D3}}$, both decreasing it (it is the case of a supersymmetric flux, including the flux on the Whitney brane) and increasing it (it is the case of a flux generating a non-zero FI-terms inducing e.g. a T-brane background \cite{Cicoli:2015ylx}). These fluxes typically do not change the order of magnitude of our estimations. However they must be taken into consideration in explicit models when computing D3-tadpole cancellation.

Cancelling the tadpole locally through (4+4) D7-branes on top of O7-planes has led to charges $[-72,8]$ in \cite{Carta:2020ohw} for CICY orientifolds and $[-60, 0]$ in \cite{Altman:2021pyc} for toric CY orientifolds with $h^{1,1}\leq 6$ from exchange involutions.\footnote{We note that $Q_{\text{D3}}=-2Q_{\text{SO}(8)}^{\text{D3}}$ in the conventions of \cite{Carta:2020ohw}, while $Q_{\text{D3}}=-2Q_{\text{D3}}^{\text{loc}}$ in the convention of \cite{Altman:2021pyc}.
Our convention for the D3-tadpole \eqref{eq:DThreeTadpoleGen} is based on eq.~(3.81) in \cite{Denef:2008wq} where $Q_{\text{D3}}=2Q_{c}$.}
In the analysis in \cite{Carta:2020ohw}, it has been shown how non-local tadpole cancellation through generic D7-branes can lead to a significantly larger range $[ -264,-24]$.
This had consciously been used in many previous applications involving Whitney branes \cite{Collinucci:2008pf,Collinucci:2008sq,Cicoli:2011qg,Crino:2020qwk}, or mild splitting of them \cite{Louis:2012nb,Braun:2015pza,Cicoli:2017shd,Cicoli:2021dhg}

We collect all divisors with Hodge numbers $h^{0,2}(D)>1$ and $h^{0,1}(D)=0$ at $h^{1,1}(X)\leq 6$ as computed in \cite{Altman:2021pyc}.
It is then instructive to compare 
the total D3-charge contribution from a stack of (4+4) D7-branes (local) and Whitney branes (non-local), as shown in Fig.~\ref{fig:ChiWD7} for $h^{1,1}(X)\leq 6$ and $7\leq h^{1,1}(X)\leq 12$ respectively, where
\begin{equation}
Q_{\text{SO}(8)}(D)=-(4+4)\cdot\dfrac{\chi(D)}{24} \kom Q_{\text{WD7}}(D)\simeq -  \dfrac{\chi(4D)}{12} - 9 \int_X D^3\:.
\end{equation}
In the last definition we have neglected the flux contribution \eqref{D3WD7flux} depending on $F-B_2$, as it does not change the order of magnitude of the Whitney brane D3-charge.

\begin{table}
\centering
\begin{minipage}{0.4\linewidth}
\centering

\textbf{Complete scan at $h^{1,1}\leq 6$}

\vspace{0.2cm}

\begin{tabular}{c|c|c}
\toprule 
$h^{1,1}(X)$ & $Q_{\text{SO}(8)}(D)$& $Q_{\text{WD7}}(D)$ \\ [0.1em]
\hline 
\hline 
 &  & \\ [-1.em]
2 & $-25.82$ & $-269.77$ \\ [0.1em]
\hline 
 &  & \\ [-1.em]
3 & $-28.31$ & $-316.69$ \\ [0.1em]
\hline 
 &  & \\ [-1.em]
4 & $-29.78$ & $-342.26$ \\ [0.1em]
\hline 
 &  & \\ [-1.em]
5 & $-30.27$ & $-344.21$ \\ [0.1em]
\hline 
 &  & \\ [-1.em]
6 & $-30.53$ & $-339.49$ \\ [0.1em]
\hline 
 &  & \\ [-1.em]
7 & $-30.16$ & $-321.89$ \\[0.1em] \bottomrule
\end{tabular} 
\end{minipage}
\hspace*{1cm}
\begin{minipage}{0.4\linewidth}
\centering

\textbf{Random data at $h^{1,1}\leq 12$}

\vspace{0.2cm}
\begin{tabular}{c|c|c}
\toprule 
$h^{1,1}(X)$ & $Q_{\text{SO}(8)}(D)$& $Q_{\text{WD7}}(D)$ \\ [0.1em]
\hline 
\hline 
 &  & \\ [-1.em]
7 & $-35.21$ & $ -391.76$ \\ [0.1em]
\hline 
 &  & \\ [-1.em]
8 & $-37.82$ & $-428.97$ \\ [0.1em]
\hline 
 &  & \\ [-1.em]
9 & $-40.12$ & $-458.18$ \\ [0.1em]
\hline 
 &  & \\ [-1.em]
10 & $-40.27$ & $-457.08$ \\ [0.1em]
\hline 
 &  & \\ [-1.em]
11 & $-42.90$ & $-497.89$ \\[0.1em]
\hline 
 &  & \\ [-1.em]
12 & $-42.23$ & $-480.87$ \\[0.1em] \bottomrule
\end{tabular} 
\end{minipage}
\caption{Average D3-charge contribution for local and non-local D7-tadpole cancellation for complete scan (left) and randomise data (right)}\label{tab:MeanD3SO8WD7} 
\end{table}

The maximal D3-charge contribution from D7-branes on a single divisor are given by
\begin{align}
|Q_{\text{SO}(8)}(D)|\, |_{\text{max}}&=\begin{cases}
183 & h^{1,1}(X)\leq 6\\
329.3 & 7\leq h^{1,1}(X)\leq 12
\end{cases}\, ,\nn\\
 |Q_{\text{WD7}}(D)|\, |_{\text{max}}&=\begin{cases}
3,585 & h^{1,1}(X)\leq 6\\
6489.3 & 7\leq h^{1,1}(X)\leq 12
\end{cases}\:.
\end{align}
We collected the average D3-charge for both sources in Tab.~\ref{tab:MeanD3SO8WD7} where $Q_{\text{WD7}}(D)$ is enhanced by a factor of $11$ on average.

\subsubsection*{Large D3-charge and genus-one fibrations}

An interesting observation concerns the behaviour of the D3-charge distribution at large $h^{1,2}$.
While one discovers no particular structure at small $h^{1,2}<100$,
the regime at large $h^{1,2}>100$ exhibits, instead of a uniform distribution, two distinct dominant lines. 
We believe that this emergent structure in the distribution of D3-charges has not yet been observed in the literature.

A hint for what is going on is obtained from previous investigations into the underlying fibration structure of toric CY threefolds at large $h^{1,2}$, see \cite{Huang:2018gpl,Huang:2018esr,Huang:2018vup,Huang:2019pne} and references therein.
It is in fact true that CY threefolds in the KS database at sufficiently large Hodge numbers ($h^{1,2}$ larger than $240$) are associated with elliptic fibrations over complex base surfaces \cite{Huang:2018gpl}.
At the level of 4D reflexive polytopes $\Delta^{\circ}$, it is quite straight forward to identify the corresponding fibrations.
Namely, whenever $\Delta^{\circ}$ contains a 2D reflexive sub-polytope, the associated CY manifold enjoys a genus one fibration \cite{Kreuzer:1997zg}.\footnote{We stress that there are some subtleties occurring when relating the fibration of the polytopes to the actual toric variety, see \cite{Huang:2019pne} for a detailed discussion.}
This is indeed a quite common feature: out of the 473.8 million polytopes listed in \cite{Kreuzer:2000xy}, only 29,223 do not contain any such 2D reflexive polytope \cite{Huang:2019pne}.\footnote{In our analysis, we encounter 2,857 (60) of these polytopes in the complete (random) data at $h^{1,1}\leq 7$ ($7\leq h^{1,1}\leq 12$).}

\begin{figure}[t!]
\centering
\includegraphics[scale=0.66]{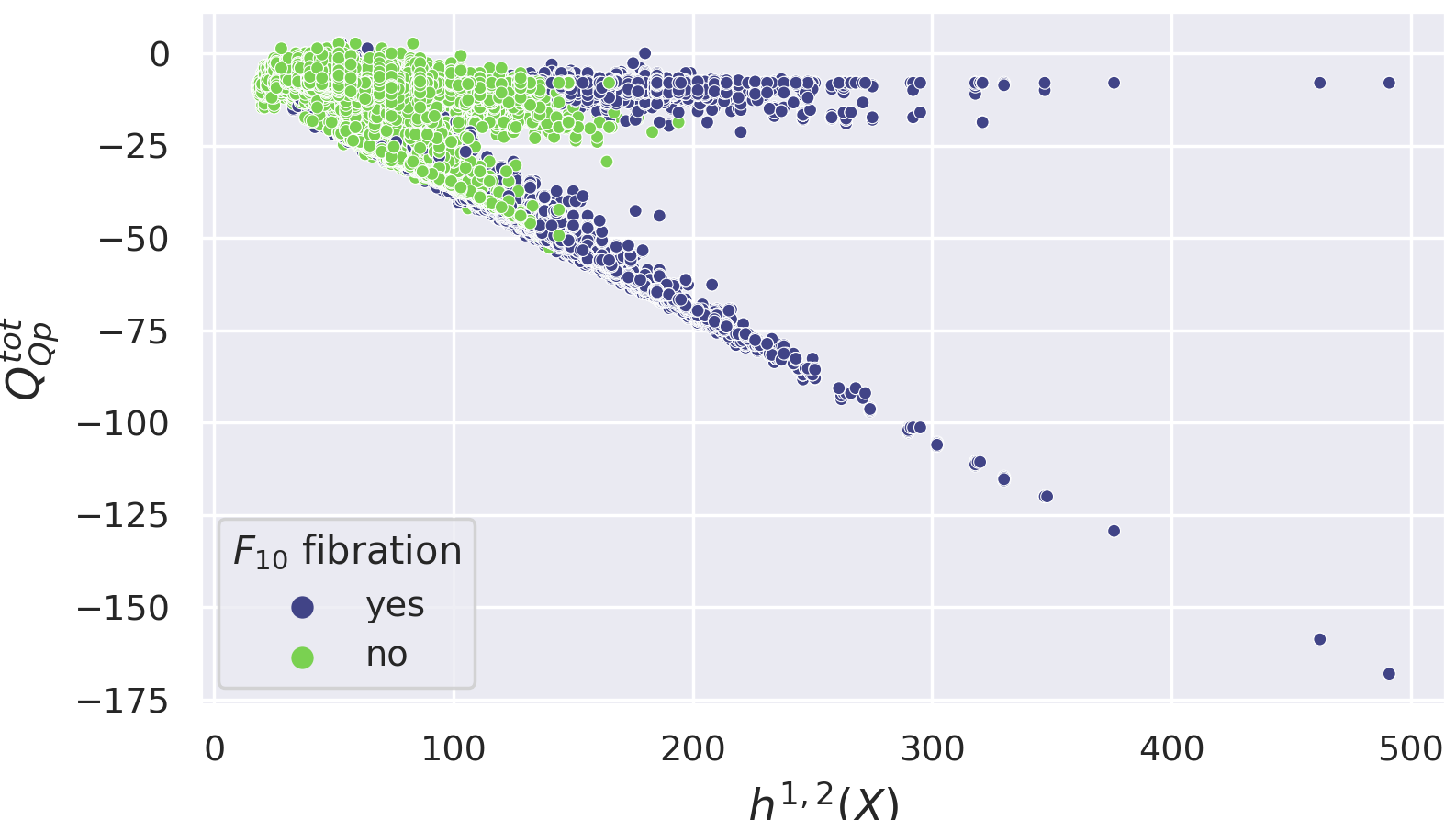}
\caption{Total D3-charge contributions from O-planes for orientifold models with colours indicating the presence of an underlying $F_{10}=\bP[2,3,1]$ fibration.}\label{fig:D3ChargesF10} 
\end{figure}

There are only 16 distinct types of genus one fibrations $F_{i}$ which can be easily identified from the classification of 2D reflexive polytopes.\footnote{A classification of the 16 distinct polytopes is provided in Appendix A of~\cite{Huang:2018vup} which were previously studied in \cite{Bouchard:2003bu} and play a role in F-theory \cite{Braun:2011ux,Braun:2013nqa,Klevers:2014bqa,Huang:2018esr}.}
At least at large Hodge numbers,
the KS database is dominated by polytopes exhibiting a description of a standard $F_{10}$ fibration \cite{Huang:2018gpl} (the elliptic fiber is a hypersurface in $\bP[2,3,1]$) which therefore also plays a distinguished role in our analysis.

Utilising the algorithm of \cite{Huang:2019pne},
we computed the 2D reflexive sub-polytopes and the fibration type for each of the favourable 4D polytopes appearing in our analysis, checking that the presence of $F_{10}$ is dominant.
We computed the D3-charge distribution for the different types of fibres. In Fig.~\ref{fig:D3ChargesF10} we report that the generic elliptic fibre $F_{10}$ dominates especially at $h^{1,2}>200$ as expected from \cite{Huang:2018gpl}.
Not surprisingly, it is responsible for the universal structure observed in Fig.~\ref{fig:OverviewD3ChargeOp} independently of $h^{1,1}(X)$.
In the regime $h^{1,2}<200$, similar sub-dominant patterns are found also for elliptic $F_{6}$ and $F_{8}$ as well as non-elliptic $F_{4}$ (the fiber is an hypersurface in $\bP^{2}[2,1,1]$) fibrations.
All other fibrations as well as the polytopes without any fibration seem not to experience any enhancement in their D3-charge contribution (i.e. they are mostly constant as functions of $h^{1,2}$) nor are they showing any particularly interesting patterns.

\

Let us try to explain what happens for the $F_{10}$ case. The CY equation takes the Weierstrass form, i.e.,
\begin{equation}\label{WeierstrassEq}
y^2=x^3 + f(w) x z^4 + g(w)  z^6\:.
\end{equation}
Here, $w$ denotes collectively coordinates on the toric two-dimensional base $B$,
whereas $x,y,z$ are projective coordinates on $\bP[2,3,1]$ with $x$ and $y$ being sections respectively of $\bar{K}_B^{\otimes 2}$ and $\bar{K}_B^{\otimes 2}$. For consistency of the equation, $f$ and $g$ must be sections respectively of $\bar{K}_B^{\otimes 4}$ and $\bar{K}_B^{\otimes 6}$

At fixed $w$, the equation \eqref{WeierstrassEq} describes a torus. The $\mathbb{Z}_2$ involution of the torus (with four fixed points) is implemented in this algebraic setup by taking $y\mapsto -y$ (or equivalently $z\mapsto-z$). The Weierstrass form is already invariant. Hence, if one takes \eqref{WeierstrassEq} as the defining equation for the CY three-fold, one has the involution that inverts $y$. This toric coordinate is manifestly of high degree (and among the coordinates of this threefold, $y$ is the highest degree one)
and correspondingly the Euler characteristic of $D_y$ is large. This is the main reason why we find the largest D3-charges for these models.

In studying the $F_{10}$ case, we realise another fact: one may add to \eqref{WeierstrassEq} also a term proportional to $x^2z^2$ and then consider the involution $x\mapsto-x$. $x$ is also high degree and the D3-charge one would obtain from such an involution is still large, even if lower than the one obtained by $y\mapsto-y$. 
However, there is a pathology: the invariant CY equation would be 
\begin{equation}
y^2=(a(w) x^2 + g(w)z^4)  z^2\:,
\end{equation}
that has a manifest (non crepantly resolvable) singularity at $z=y=0$. Since $xyz$ is the SR-ideal, the D7/O7's do not touch the singularity and their topological invariants do not feel the pathology. However, we excluded it from our analysis as $\Delta_{k}$ is not reflexive because the monomial $x^{3}$ is associated with a vertex in the full dual polytope $\Delta$.
If we had included such models, we would have obtained a second diagonal line in our plots of models with large D3-charge.

\section{Example with $(h^{1,1},h^{1,2})=(11,491)$}\label{sec:H11_11Example}

\begin{table}[t!]
\centering
\begin{tabular}{|c|c|c|c|c|c|c|c|c|c|c|c|c|c|c||c|}
\hline 
 &  &  &  &  &  &  & &&   &&  & &&  &   \\ [-1.em]
$z_{1}$ & $z_{2}$ & $z_{3}$ & $z_{4}$ & $z_{5}$ & $z_{6}$ & $z_{7}$ & $z_{8}$ &$z_{9}$ & $z_{10}$ & $z_{11}$ &$z_{12}$ &$z_{13}$ & $z_{14}$ &$z_{15}$ &   \\ [0.3em]
\hline 
\hline 
 &  &  &  &  &  &  &&  &  &&  & &&  &  \\ [-0.9em]
 1& 1 & 12 & 28 & 42 & 0 & 0 & 0 & 0 & 0& 0 & 0 & 0& 0 & 0 & 84 \\ [0.2em]
\hline 
 &  &  &  &  &  &  &&  &  &&  & &&  &  \\ [-0.9em]
0 & 0 & 6 & 14 & 21 & 1 & 0 & 0 & 0 & 0& 0 & 0 & 0& 0 & 0 & 42 \\ [0.2em]
\hline 
 &  &  &  &  &  &  &&  &  &&  & &&  &  \\ [-0.9em]
0 & 0 & 5 & 12 & 18 & 0 & 1 & 0 & 0 & 0& 0 & 0 & 0& 0 & 0 & 36 \\ [0.2em]
\hline 
 &  &  &  &  &  &  &&  &  &&  & &&  &  \\ [-0.9em]
0 & 0 & 4 & 10 & 15 & 0 & 0 & 1 & 0 & 0& 0 & 0 & 0& 0 & 0 & 30 \\ [0.2em]
\hline 
 &  &  &  &  &  &  &&  &  &&  & &&  &  \\ [-0.9em]
0 & 0 & 4 & 9 & 14 & 0 & 0 & 0 & 1 & 0& 0 & 0 & 0& 0 & 0 & 28 \\ [0.2em]
\hline 
 &  &  &  &  &  &  &&  &  &&  & &&  &  \\ [-0.9em]
0 & 0 & 3 & 8 & 12 & 0 & 0 & 0 & 0 & 1& 0 & 0 & 0& 0 & 0 & 24 \\ [0.2em]
\hline 
 &  &  &  &  &  &  &&  &  &&  & &&  &  \\ [-0.9em]
0 & 0 & 3 & 7 & 10 & 0 & 0 & 0 & 0 & 0& 1 & 0 & 0& 0 & 0 & 21 \\ [0.2em]
\hline 
 &  &  &  &  &  &  &&  &  &&  & &&  &  \\ [-0.9em]
0 & 0 & 2 & 6 & 9 & 0 & 0 & 0 & 0 & 0& 0 & 1 & 0& 0 & 0 & 18 \\ [0.2em]
\hline 
 &  &  &  &  &  &  &&  &  &&  & &&  &  \\ [-0.9em]
0& 0 & 2 & 4 & 7 & 0 & 0 & 0 & 0 & 0& 0 & 0 & 1& 0 & 0 & 14 \\ [0.2em]
\hline 
 &  &  &  &  &  &  &&  &  &&  & &&  &  \\ [-0.9em]
0 & 0 & 1 & 4 & 6 & 0 & 0 & 0 & 0 & 0& 0 & 0 & 0& 1 & 0 & 12 \\ [0.2em]
\hline 
 &  &  &  &  &  &  &&  &  &&  & &&  &  \\ [-0.9em]
0 & 0 & 0 & 2 & 3 & 0 & 0 & 0 & 0 & 0& 0 & 0 & 0& 0 & 1 & 6 \\ [0.2em]
\hline 
\end{tabular} 
\caption{Weights for the model with $h^{1,1}=11$ and $h^{1,2}=491$.}\label{tab:WeightsExample5THOO11} 
\end{table}

To be more specific,
let us describe in more detail the model with the potentially largest D3-tadpole reported in Tab~\ref{tab:D3charge}.
It turns out that this model is obtained from an involution of a CY threefold $X$ with Euler characteristic $\chi(X)=-960$ and Hodge numbers $(h^{1,1},h^{1,2})=(11,491)$.
The GLSM charges of $X$ are collected in Tab.~\ref{tab:WeightsExample5THOO11};
the SR ideal is given by
\begin{align}\label{eq:SRIdealModelH11_11} 
I_{\text{SR}}&=\lbrace z_{1}z_{2},z_{3}z_{6},z_{3}z_{7},z_{3}z_{8},z_{3}z_{9},z_{3}z_{10},z_{3}z_{11},z_{3}z_{12},z_{3}z_{13},z_{3}z_{14},z_{4}z_{6},z_{4}z_{7},z_{4}z_{8},z_{4}z_{9},  \nn\\
&\quad z_{4}z_{10},z_{4}z_{11},z_{4}z_{12},z_{4}z_{14},z_{5}z_{6},z_{5}z_{7},z_{5}z_{8},z_{5}z_{9},z_{5}z_{10},z_{5}z_{12},z_{6}z_{8},z_{6}z_{10},z_{6}z_{12},z_{6}z_{13},\nn\\
&\quad z_{6}z_{14},z_{6}z_{15}, z_{7}z_{10},z_{7}z_{12},z_{7}z_{13},z_{7}z_{14},z_{7}z_{15},z_{8}z_{12},z_{8}z_{13},z_{8}z_{14},z_{8}z_{15},z_{9}z_{12},z_{9}z_{14},\nn\\
&\quad z_{9}z_{15},z_{10}z_{14},z_{10}z_{15},z_{11}z_{15},z_{12}z_{15},z_{4}z_{5}z_{15},z_{5}z_{13}z_{14},z_{5}z_{13}z_{15},z_{7}z_{9}z_{11},z_{8}z_{9}z_{11},\nn\\
&\quad z_{9}z_{10}z_{11},z_{10}z_{11}z_{13},z_{11}z_{12}z_{13},z_{11}z_{13}z_{14}\rbrace\, .
\end{align}
The 2nd Chern numbers are:
\begin{equation}
\int_{D_{i}}\, c_{2}(X)=\lbrace 24  ,24 ,168 ,368, 548 , -4  ,-4,  -4 , -4 , -4  ,-4 , -4 , -4  ,-4,  -4\rbrace\, .
\end{equation}
Finally, the Hodge numbers of the divisors can be computed to be:
\begin{align}\label{eq:HodgeNumExH11_11} 
h^{\bullet}(D_{1})=h^{\bullet}(D_{2})&=\lbrace 1,0,1,20 \rbrace\kom  h^{\bullet}(D_{3})=\lbrace 1,0,13,140 \rbrace\, ,\nn\\
h^{\bullet}(D_{4})&=\lbrace 1,0,51,392 \rbrace\kom h^{\bullet}(D_{5})=\lbrace 1,0,118,750 \rbrace\, ,\nn\\
h^{\bullet}(D_{i})&=\lbrace 1,0,0,2 \rbrace\kom i=6,\ldots ,15\, .
\end{align}

Related to the discussion above, one finds that this CY exhibits an $F_{10}$ fibration with coordinates $z_{4},z_{5},z_{15}=x,y,z$ over the Hirzebruch surface $\mathbb{F}_{12}$ as can be seen from the last line in the GLSM charge matrix in Tab.~\ref{tab:WeightsExample5THOO11}.\footnote{In fact, $h^{1,2}=491$ is the largest possible value for any elliptic CY threefold \cite{Taylor:2012dr}.}
Our analysis shows that the allowed values of the D3-charge from O$p$-planes are $8\leq |Q^{\text{tot}}_{\text{O}p}|\leq 168$.
The maximally allowed D3-charge from O7-planes is actually obtained from (recall \eqref{eq:D3OpHodgeNum} and that all other $D_{i>5}$ are $\mathrm{dP}_{1}$ divisors)
\begin{equation}
\chi(D_{5})+\chi(D_{15})+4\cdot \chi(\mathrm{dP}_{1})=2(h^{1,2}+h^{1,1}+2)=1,008\, .
\end{equation}
It is associated with the standard involution of the torus fibre $z_{5}\raw -z_{5}$ as argued above.

\

For this reason, let us study this involution
\begin{equation}
z_{5}\rightarrow -z_{5}
\end{equation}
which gives rise to O7-planes on $D_{5}$, $D_{6}$, $D_{8}$, $D_{12}$, $D_{13}$ and $D_{15}$ and invariant Hodge numbers $(h^{1,2}_{-},h^{1,2}_{+})=(491,0)$. There are no O3-planes.
As it can be read from the Hodge numbers \eqref{eq:HodgeNumExH11_11}, the Euler characteristic of the O7 divisors are $\chi(D_5)=988$ and $\chi(D_i)=4$ for $i=6,8,12,13,15$. Hence, 
The O7-planes contribute to the D3 charge with: 
\begin{equation}
Q_{\text{O}7}^{\text{tot}}=-\sum_{k=5,6,8,12,13,15} \frac{\chi(D_k)}{6}= -168\: .
\end{equation}

As concerns the branes configuration, the divisors  $D_{6}$, $D_{8}$, $D_{12}$, $D_{13}$ and $D_{15}$ are rigid and then  support an $\mathrm{SO}(8)$ stack.\footnote{The $\mathrm{SO}(8)$ stacks do not intersect each other.} The D7-tadpole from the $D_5$ divisors will instead be canceled by a Whitney brane.

We choose a B-field that allow to have zero flux on each D7-brane:
\begin{equation}\label{eq:B2h11bis} 
	B_2=\frac{1}{2}\left(D_6+D_8+D_{12}+D_{13}+D_{15}\right)\:.
\end{equation}
Since the divisors $D_{6,8,12,13,15}$ do not intersect each other, the pull-back of the B-field on the divisor $D_i$ is equal to $\iota_{D_i}^* B_2 = \frac{D_i}{2}$ and then it cancels the non-integral flux that is necessary for Freed-Witten anomaly cancellation, leading to $\mathcal{F}_i=0$.  As regarding the Whitney brane, we need to check that there exists an integral 2-form $F$ that cancels either $\frac32 D_5+B_2$ or $\frac32 D_5-B_2$ in \eqref{LineBndlsFluxWhit}. 
This happens, because $D_5+B_2$ is an even form, as it can be checked rom the GLSM weights in Table~\ref{tab:WeightsExample5THOO11}.

Taking vanishing fluxes on each D7-brane, the D3-charge contribution is only geometrical. 
The $\mathrm{SO}(8)$ stacks contribute to the D3-charge as
\begin{equation}
 Q_{\mathrm{SO}(8)}^{\text{tot}} = -\sum_{i=6,8,12,13,15} \frac{\chi(D_i)}{3} = - 5\cdot \frac43 = - \frac{20}{3}  \,,\qquad i=6,8,12,13,15\:,
\end{equation} 
while the main contribution to the D3-charge comes from the Whitney brane, whose 
geometric contribution \eqref{D3WD7geom} is 
\begin{equation}
	Q_{WD7,{\rm geom}} = -\frac{\chi(4D_5)}{12} - 9 \int_X D_5^3   =-\frac{19,468}{3} \;,
\end{equation}
where we used $\chi(4D_{5})=30,352$ and $D_5^3=440$.
Cancelling the D7-tadpole from $D_5$ by a Whitney brane, instead of an $\mathrm{SO}(8)$ stack, increases the D3-charge from 7-branes by approximately a factor of
\begin{equation}
\dfrac{Q_{WD7,{\rm geom}}^{D_5} }{Q_{\mathrm{SO}(8)}^{D_5}}\approx 20\, ,
\end{equation}
where
\begin{equation}
Q_{\mathrm{SO}(8)}^{D_5}=-\frac{\chi(D_5)}{3}=-\frac{988}{3}\, .
\end{equation}
The total D3-charge contribution from localised sources is then
\begin{eqnarray}
Q_{\text{D3}}  &=& Q_{\text{O}7}^{\text{tot}} +  Q_{\mathrm{SO}(8)}^{\text{tot}} + Q_{WD7,{\rm geom}} = -6664\:,
\end{eqnarray}
as reported in Table~\ref{tab:D3charge}.

\

To stabilise all the moduli via non-perturbative effects,
it would be favourable to have instantons on the other rigid divisors.
Since the B-field \eqref{eq:B2h11bis} does not allow to have vanishing fluxes $\mathcal{F}_{E3}$ on any of these divisors we cannot have O(1) instantons.
On the other side, rank-2 instantons might be allowed \cite{Berglund:2012gr} provided that one checks that no chiral modes arise at the intersection with the $\mathrm{SO}(8)$ stacks.
This model is of course not suitable for anti-D3 uplift since there are no O3-planes, but in principle we could engineer a T-brane background that allows for de Sitter minima \cite{Cicoli:2015ylx}.

\section{Conclusions}\label{sec:Conclusions} 

In this paper,
we generated a database of CY orientifolds from holomorphic reflection involutions of CY hypersurfaces. We determined the orientifold configurations for all favourable FRSTs for 
$h^{1,1}\leq 7$.
 We found more than 70 million involutions of which over 20 million correspond to smooth compactifications. Singular involutions were identified and their structure deserves further investigation. We also specified the number of cases with either O3 or O7 planes suitable for antibrane or T-brane uplifts. 
 
We plotted several relevant quantities such as the Euler number and Hodge numbers of the divisors and the value of the D3 brane charges. We observed some interesting patterns in the distribution of the models. In particular the values of the D3 charges show non-trivial structures, such as higher concentration of models in some particular directions, that would be interesting to understand from the more mathematical perspective.

Our algorithm is in principle capable of computing orientifolds for any $h^{1,1}$.
 We provided partial results for triangulations up to $h^{1,1}=12$.
We found several classes of models with different behaviour in their D3-charge and O-plane configuration.
Most importantly,
we provided evidence for a  large class of models for which the D3-charge from O$p$-planes grows $\sim -(h^{1,1}_{+}+h^{1,2}_{-})/3$, i.e., linearly with the number of invariant geometric moduli.
This constitutes an upper bound on the absolute value of the total D3-charge from D7/O7's and O3's.

We further showed that cancelling the D7-tadpole non-locally via Whitney branes as opposed to locally via $\mathrm{SO}(8)$ stacks on top of O7-planes increases the overall D3-charge by up to factors of $12$.
We presented an explicit orientifold with Hodge numbers $(h^{1,1},h^{1,2})=(11,491)=(h^{1,1}_{+},h^{1,2}_{-})$ which led to a total D3-charge of $|Q_{\text{D3}}|=6,664$.
This value beats previous D3-charge records in type IIB by a large margin (recall Tab.~\ref{tab:OverviewD3ChargesLit}).
It provides the necessary space to turn on background fluxes which in turn are relevant for stabilising moduli and model building.
Beyond that,
our database contains a plethora of other models, $357,730$ to be precise, with $|Q_{\text{D3}}|> 504$
making it an excellent starting point for the construction of trustable string vacua.
An explicit calculation of moduli stabilisation for these vacua is beyond the scope of this paper.

An important result of this paper concerns the non-trivial D3-charge distribution as a function of $h^{1,2}$.
We provided evidence based on the existence of 2D reflexive sub-polytopes that this is mainly a result of special genus one fibrations of the associated CY threefolds, especially elliptic $F_{10}$ (hypersurface in $\bP[2,3,1]$) and non-elliptic $F_{4}$ (hypersurface in $\bP[1,1,2]$) fibrations.
The patterns observed in Fig.~\ref{fig:OverviewD3ChargeOp} are directly linked to reflecting either coordinates along fibre or the base.
Further, we put forward an argument for $F_{10}$ fibrations that involutions involving coordinates along the fibre generically maximise the bound on the D3-charge.
It would be interesting to further explore the role of genus one fibrations in the context of $\cN=1$ compactifications of type IIB to 4 dimensions.

In the future,
it is desirable to extend the database in the regime $h^{1,1}\geq 12$.
Recent works \cite{Long:2014fba,Demirtas:2018akl,Demirtas:2020dbm} demonstrated that triangulations of polytopes with large $h^{1,1}$ can be constructed efficiently.
However, exhaustive scans or random sampling might be impractical which is why a more targeted approach by employing optimisation methods would be favourable as previously applied in the search for string vacua \cite{Blaback:2013ht,Blaback:2013fca,Abel:2014xta,Cole:2019enn,Larfors:2020ugo,AbdusSalam:2020ywo,Bena:2021wyr,Krippendorf:2021uxu}.
In the same spirit, it would also be exciting to relate our database to the one of CICYs \cite{Carta:2020ohw} and combine it with the one for divisor exchange involutions \cite{Altman:2021pyc}.
For instance, as compared to \citep{Altman:2021pyc}, we have not glued together the Kähler cones of equivalent triangulations.
Similar to \cite{Carta:2020ohw},
a large fraction of the orientifolds contained in the database are singular which can in special cases like the conifold be resolved as discussed in \cite{Carta:2021uwv} for the CICY landscape.
Such resolutions might lead to new CY threefolds that are not contained in the KS database.

\acknowledgments
We would like to thank Andres Collinucci, Xin Gao, Arthur Hebecker, Sven Krippendorf, Francesco Muia and Pramod Shukla for useful discussions.
AS acknowledges support by the German Academic Scholarship Foundation and by DAMTP through an STFC studentship. The work of FQ has been partially supported by STFC consolidated grants ST/P000681/1, ST/T000694/1. C.C. and R.V. acknowledge support by INFN Iniziativa Specifica ST\&FI.

\appendix

\section{Examples with Whitney branes}\label{app:EWB} 

In this appendix,
we study two CYs at $h^{1,1}=3$ which admit divisors of different topologies.
For each toric divisor $D_i$ we study a Whitney brane given by the equation
\begin{equation}\label{eq:WhBrDef} 
\eta_{i}^{2}-z_{i}^{2}\chi_{i}=0\, ,
\end{equation}
in order to see whether the line bundles  $\mathcal{O}(4D_i)$ and $\mathcal{O}(6D_i)$ force the locus \eqref{eq:WhBrDef} to factorise.

All in all, our analysis suggests that $h^{0,2}(D)>1$ always leads to proper Whitney branes, while for divisors with $h^{0,2}(D)=1$ the factorisation depends on the actual GLSM weight matrix.
In any case, we are mostly interested in divisors of maximal Euler number for which generically $h^{0,2},h^{1,1}\gg 1$.

\subsection{Example with an $\mathrm{SO}(8)$ stack for a non-rigid SD1 divisor}

\begin{table}[t!]
\centering
\begin{tabular}{|c|c|c|c|c|c|c||c|}
\hline 
 &  &  &  &  &  &  &  \\[-0.8em] 
$z_{1}$ & $z_{2}$ & $z_{3}$ & $z_{4}$ & $z_{5}$ & $z_{6}$ & $z_{7}$ & $D_{H}$ \\ [0.3em]
\hline 
\hline 
 &  &  &  &  &  &  &  \\[-0.8em] 
0 & 0 & 1 & 1 & 1 & 1 & 4 & 8 \\ [0.3em]
\hline 
 &  &  &  &  &  &  &  \\[-0.8em] 
0 & 1 & 0 & 0 & 1 & 1 & 3 & 6 \\ [0.3em]
\hline 
 &  &  &  &  &  &  &  \\[-0.8em] 
1 & 0 & 1 & 0 & 1 & 2 & 5 & 10 \\ [0.3em]
\hline 
\end{tabular} 
\caption{Weights for polytope with ID $237$ at $h^{1,1}=3$.}\label{tab:weightsH113Mod4} 
\end{table}


We consider the model (\texttt{POLYID}: 237, \texttt{TRIANGN}: 1 in \cite{Altman:2014bfa}) with weight matrix in Tab.~\ref{tab:weightsH113Mod4} and SR ideal
\begin{equation}
I_{\text{SR}}=\lbrace z_{1} z_{6} , z_{2}  z_{5} , z_{3}  z_{4}  z_{7}  \rbrace\, .
\end{equation}
Following the procedure outlined in Sect.~\ref{sec:DivTops},
we computed the Hodge numbers
\begin{align}
h^{\bullet}(D_{1})&=\lbrace 1,0,0,9 \rbrace\kom h^{\bullet}(D_{2})=\lbrace 1,0,0,8 \rbrace\kom h^{\bullet}(D_{3})=\lbrace 1,0,1,21 \rbrace\, ,\nn\\
h^{\bullet}(D_{4})&=\lbrace 1,0,0,12 \rbrace\kom h^{\bullet}(D_{5})=\lbrace 1,0,2,29 \rbrace\kom h^{\bullet}(D_{6})=\lbrace 1,0,3,38 \rbrace\, ,\nn\\
h^{\bullet}(D_{7})&=\lbrace 1,0,26,177 \rbrace\, .
\end{align}
We have three rigid divisors $D_{1}, D_{2}, D_{4}$ with $D_{1}$ a $\mathrm{dP}_{8}$ and $D_{2}$ a $\mathrm{dP}_{7}$,
one SD1 divisor $D_{3}$ and three non-rigid (deformation) divisors $D_{5}, D_{6}, D_{7}$.

Let us now study the defining equation \eqref{eq:WhBrDef} for D7-brane configurations on each of the divisors.
For the rigid divisors $D_{1}, D_{2}, D_{4}$, the generic section of  $\mathcal{O}(4D_i),\mathcal{O}(6D_i)$ are forced to factorise as
\begin{equation}
\eta_{i}=z_{i}^{4}\kom \chi_{i}=z_{i}^{6}\quad ,\, i\in\lbrace 1,2,4\rbrace\, .
\end{equation}
giving an $\mathrm{SO}(8)$ stack.
In contrast, we have generic polynomials for the non-rigid divisors $D_{5}, D_{6}, D_{7}$ and hence proper Whitney brane configurations.

The more interesting scenario concerns the SD1 divisor $D_{3}=\lbrace z_{3}=0\rbrace$.
Looking at the GLSM charges in Tab.~\ref{tab:weightsH113Mod4}, the degrees for $z_{3}$ are given by $(1,0,1)$ which implies that $z_{3}=0$ can be modified only through combinations of $z_{1}$ and $z_{4}$ with weights $(0,0,1)$ and $(1,0,0)$ respectively.
This is because all other coordinates $z_{i}$, $i\neq 1,3,4$, have degrees $(*,1,*)$.
Thus, we may equivalently write
\begin{equation}
z_{3}+\alpha z_{1}z_{4}=0
\end{equation}
which is the only possible deformation of $D_{3}$ and hence $h^{0,2}(D_{3})=1$.

The Whitney brane is a representative of the class $8[D_{3}]$ with degrees $(8,0,8)$. A generic element of this class is of the form
\begin{equation}
P_8(z_3\,,\, z_1z_4) \equiv \sum_{i=0}^8\, \alpha_{i}\, z_{3}^{i}\, (z_{1}z_{4})^{8-i}=0\, ,
\end{equation}
where $P_{8}$ is a homogeneous polynomial of degree $8$ in two variables.
Clearly, the  equation $P_8(X,Y)=0$ admits precisely $8$ zeros which allows us to write it as
\begin{equation}\label{eq:Prod4p4model1}
\prod_{i=1}^{4}\, (z_{3}-\beta_{i}(z_{1}z_{4}))(z_{3}+\beta_{i}(z_{1}z_{4}))=0\, ,
\end{equation}
where we also imposed that our representative is an invariant locus under the involution $z_3\mapsto-z_3$. This generic factorisation is valid for all invariant representatives of $8[D_3]$, hence also for the Whitney brane in this class.

The equation \eqref{eq:Prod4p4model1} tells us that the Whitney brane corresponding to the divisor $D_3$ is forced to factorise into 4 pairs of brane/image-brane, that need not necessarily be parallel, i.e., they can in principle intersect.\footnote{For K3 divisors, we expect to find similar situations where the D7-branes are however expected to be parallel without any intersection.}

Notice that the above argument would fail if there was an additional coordinate $z_{0}$ with degrees $(2,0,1)$ for which e.g. the class $2[D_{3}]$ is represented by
\begin{equation}
z_{3}^{2}+z_{3}z_{1}z_{4}+(z_{1}z_{4})^{2}+z_{0}z_{1}=0\, .
\end{equation}
The additional monomial $z_{0}z_{1}$ spoils the factorisation of the branes discussed above.
We see no reason for why such situations should not be realised in the KS database.
Indeed,
the next section provides an explicit example with a divisor with $h^{0,2}=1$ that looks topologically like a K3 divisor, but whose Whitney brane does not factorise.

\subsection{Example with a divisor with $h^{p,q}=h^{p,q}(K3)$}

\begin{table}[t!]
\centering
\begin{tabular}{|c|c|c|c|c|c|c||c|}
\hline 
 &  &  &  &  &  &  &  \\[-0.8em] 
$z_{1}$ & $z_{2}$ & $z_{3}$ & $z_{4}$ & $z_{5}$ & $z_{6}$ & $z_{7}$ & $D_{H}$ \\ [0.3em]
\hline 
\hline 
 &  &  &  &  &  &  &  \\[-0.8em] 
0 & 0 & 1 & 1 & 1 & 2 & 1 & 6 \\ [0.3em]
\hline 
 &  &  &  &  &  &  &  \\[-0.8em] 
0 & 1 & 0 & 0 & 1 & 1 & 0 & 3 \\ [0.3em]
\hline 
 &  &  &  &  &  &  &  \\[-0.8em] 
1 & 0 & 2 & 3 & 2 & 4 & 0 & 12 \\ [0.3em]
\hline 
\end{tabular} 
\caption{Weights for polytope with ID $57$ at $h^{1,1}=3$.}\label{tab:weightsH113Mod11} 
\end{table}


We consider the model (\texttt{POLYID}: 57, \texttt{TRIANGN}: 3 in \cite{Altman:2014bfa}) with weight matrix in Tab.~\ref{tab:weightsH113Mod11} and SR ideal
\begin{equation}
I_{\text{SR}}=\lbrace z_{1} z_{4} , z_{2}  z_{5} , z_{3}  z_{6}  z_{7}  \rbrace\, .
\end{equation}
We find that the Hodge numbers for the toric divisors are given by
\begin{align}
h^{\bullet}(D_{1})&=\lbrace 1,0,0,10 \rbrace\kom h^{\bullet}(D_{2})=\lbrace 1,0,0,8 \rbrace\kom h^{\bullet}(D_{3})=\lbrace 1,0,1,20 \rbrace\, ,\nn\\
h^{\bullet}(D_{4})&=\lbrace 1,0,2,30 \rbrace\kom h^{\bullet}(D_{5})=\lbrace 1,0,2,28 \rbrace\kom h^{\bullet}(D_{6})=\lbrace 1,0,6,56 \rbrace\, ,\nn\\
h^{\bullet}(D_{7})&=\lbrace 1,1,0,2 \rbrace\, .
\end{align}
We have two rigid divisors $D_{1}, D_{2}$ with $D_{2}$ a $\mathrm{dP}_{7}$, one Wilson divisor $D_{7}$, one SD2 divisor $D_{4}$ and two additional non-rigid (deformation) divisors $D_{5}, D_{6}$.
The last divisor $D_{3}$ looks topologically like a K3 surface.
Below we argue why it is not actually the case.

For the rigid divisors $D_{1}, D_{2}$ and the Wilson divisor $D_{7}$, we have $\mathrm{SO}(8)$ stacks.
For the non-rigid divisors $D_{4},D_{5}, D_{6}$, we have generic polynomials and hence proper Whitney brane configurations.

For the would-be K3 divisor $D_{3}$, a closer inspection of the weight system in Tab.~\ref{tab:weightsH113Mod11} shows that the equation $z_{3}=0$ can be deformed such that
\begin{equation}
z_{3}+\alpha z_{1}^{2}z_{7}=0
\end{equation}
and, given that this is the only possible deformation, $h^{0,2}(D_{3})=1$.
On the other hand,
the class $2[D_{3}]$ may be represented by
\begin{equation}
z_{3}^{2}+\alpha_{1}z_{3} z_{1}^{2}z_{7}+\alpha_{2} (z_{1}^{2}z_{7})^{2}+\beta z_{1}z_{4}z_{7}=0\, .
\end{equation}
This implies that $z_{3}^{8}=0$ can be modified in such a way that
\begin{equation}
\sum_{i=0}^{8}\sum_{j=0}^{4}\, \alpha_{ij}z_{3}^{8-i-2j} (z_{1}^{2}z_{7})^{i}\, (z_{1}z_{4}z_{7})^{j}=0\, .
\end{equation}
This is a non-homogeneous polynomial in the three coordinates $z_{3}$, $z_{1}^{2}z_{7}$ and $z_{1}z_{4}z_{7}$.
In particular, it does not factorise which suggests that we obtain a fully recombined D7-brane in the class $8[D_{3}]$.

We now argue that the above obstruction to the factorisation of the Whitney brane appears because $D_{3}$ is not a  K3 surface.
In fact, a K3 sourface has trivial first Chern class $c_1(K3)$. If it is embedded as a divisor $S$ into a CY threefold, $c_1(S)=-\iota_S^*S$, then
\begin{equation}
\iota_S^*S = 0  \qquad\Rightarrow\qquad \int_X S\wedge S \wedge D=0 \,\,\,\forall D\in H^{1,1}(X)
\end{equation}
The Hodge numbers are basically determined (when $h^{1,0}=0$) by the Euler characteristic and arithmetic genus of $S$, that only depend on (see \eqref{eq:EulerDiv}, \eqref{eq:AGDiv}) $\int_X S^3$ and $\int_X S^2\cdot c_2(X)$. 

In our example, $\int_X D_{3}^{3}=0$ and $\int_X D_{3}^{2}c_2(X)=24$ (and $h^{1,0}(D_3)=0$), hence giving the Hodge numbers of a K3. However, 
\begin{equation}
\int_X D_{3}\wedge D_{3}\wedge D_{i}=\kappa_{33i}=2\kom i=2,5,6\, .
\end{equation}

The above situation seems to be quite generic and happens for several other examples such as in the polytopes (triangulations) with IDs 193 (3), 60 (1), 205 (6), 247 (2) and 57 (2) in the database of \cite{Altman:2014bfa}.

\section{Simple example of CY with genus one fibrations: $\mathbb{P}[1,1,1,6,9]$}\label{app:GenusOneFib}

\begin{table}[t!]
\centering
\begin{tabular}{|c|c|c|c|c|c||c|}
\hline 
 &  &  &  &  &  &    \\[-0.8em] 
$z_{1}$ & $z_{2}$ & $z_{3}$ & $z_{4}$ & $z_{5}$ & $z_{6}$  & $D_{H}$ \\ [0.3em]
\hline 
\hline 
 &  &  &  &  &  &    \\[-0.8em] 
1 & 1 & 1 & 6 & 9 & 0  & 18 \\ [0.3em]
\hline 
 &  &  &  &  &    &  \\[-0.8em] 
0 & 0 & 0 & 2 & 3 & 1 & 6 \\ [0.3em]
\hline 
\end{tabular} 
\caption{Weights for $\mathbb{P}[1,1,1,6,9]$.}\label{tab:weightsH112} 
\end{table}

Let us show an established example with a fibration, namely the degree 18 hypersurface in $\mathbb{P}[1,1,1,6,9]$ \cite{Candelas:1994hw,Diaconescu:1999vp,Denef:2004dm} which is also prominently featured in the LVS \cite{Balasubramanian:2005zx}.
It corresponds to an elliptic fibration over $\bP^{2}$ with fibres $F_{10}$ (hypersurface in $\bP[2,3,1]$) and weights summarised in Tab.~\ref{tab:weightsH112}.
The SR-ideal reads
\begin{equation}
I_{\text{SR}}=\lbrace z_{1}z_{2}z_{3},z_{4}z_{5}z_{6}\rbrace
\end{equation}
and the topology of divisors is
\begin{align}
h^{\bullet}(D_{i})&=\lbrace 1,0,2,30 \rbrace\kom \chi(D_{i})=36\kom i=1,2,3\, ,\\
h^{\bullet}(D_{4})&=\lbrace 1,0,28,218 \rbrace\kom \chi(D_{4})=276\, ,\\
h^{\bullet}(D_{5})&=\lbrace 1,0,65,417 \rbrace\kom \chi(D_{5})=549\, ,\\
h^{\bullet}(D_{6})&=\lbrace 1,0,0,1 \rbrace\kom \chi(D_{6})=3\, .
\end{align}
This CY threefold $X_{3}$ has Hodge numbers $(h^{1,1},h^{1,2})=(2,272)$ and Euler characteristic $\chi(X)=-540$.
The most general CY equation with degrees $D_H$ in Table~\ref{tab:weightsH112} reads
\begin{align}
z_{5}^{2}&=z_{4}^{3}+h_{18}(z_{1},z_{2},z_{3})\, z_{5}z_{6}^{3}+h_{12}(z_{1},z_{2},z_{3}) z_{4}z_{6}^{4}+h_{18}(z_{1},z_{2},z_{3}) z_{6}^{6}\nn\\
&\quad +h_{3}(z_{1},z_{2},z_{3}) z_{4}z_{5}z_{6}+h_{6}(z_{1},z_{2},z_{3})  z_{4}^{2}z_{6}^{2}\, ,
\end{align}
that, by a coordinate change can be brought in a Weierstrass form.
Let us denote the $\bP^{2}$ base of $X_{3}$ as $B$ and the associated canonical class as $K_{B}$.
Then $h_{3n}$ are sections of $\cO(-nK_{B})$.

In the notation of \cite{Candelas:1994hw}, we may write $D_{6}=H-3L$ where $D_{4}=2H$, $D_{5}=3H$ and $D_{i}=L$, $i=1,2,3$.
The intersection pattern is
\begin{equation}
L^{3}=0\kom L^2H=1    \kom L\,H^2=3\kom H^{3}=9\:. 
\end{equation}
From $ c_{2}(X_{3})\cdot L=36$ and $c_{2}(X_{3})\cdot H=102$ we compute
\begin{equation}
\chi(L)=36\kom \chi(H)=111\kom \chi(H-3L)=\chi(H)-\chi(3L) \, .
\end{equation}

This example is a good arena to understand the emergence of the three lines persisting at large $h^{1,2}>100$ independently of $h^{1,1}$ as shown in Fig.~\ref{fig:OverviewD3ChargeOp}.
The orientifolds obtained from reflection involutions of one of the coordinates $z_{i}$, $i=1,\ldots ,6$, fall precisely in three categories.
A simple analysis shows that the O-plane configurations are given by:
\begin{itemize}
\item $z_{i}\raw -z_{i}$, $i=1,2,3$: A single O7-plane wrapping $D_{i}$, one O3-plane at $z_{j}=z_{k}=z_{6}=0$ and three coinciding O3-planes at $z_{j}=z_{k}=z_{5}=0$ where $(i,j,k)\in \lbrace (1,2,3), (2,3,1), (3,1,2)\rbrace$.
The Hodge numbers are $(h^{1,2}_{+},h^{1,2}_{-})=(128,144)$.
\item $z_{4}\raw -z_{4}$: A single O7-plane wrapping $D_{4}$ and Hodge numbers $(h^{1,2}_{+},h^{1,2}_{-})=(69,203)$.
\item $z_{i}\raw -z_{i}$, $i=5,6$: Two O7-planes wrapping both $D_{5}, D_{6}$ and Hodge numbers $(h^{1,2}_{+},h^{1,2}_{-})=(0,272)$.
\end{itemize}
The D3-charges from O$p$-planes are computed as
\begin{equation}
Q^{\text{tot}}_{\text{O}p}=
\begin{cases}
-8& \text{reflecting } z_1,z_{2},z_{3}  \, ,\\
-46& \text{reflecting } z_4 \, ,\\
-92 & \text{reflecting } z_5, z_6 \ .
\end{cases}
\end{equation}
Reflecting along the base $\bP^{2}$ described by $\lbrace z_1, z_2, z_3\rbrace$ gives the minimal D3-charge contribution.
The $\bP[2,3,1]$-fibre is parametrised by $\lbrace z_4, z_5, z_6\rbrace$ for which we distinguish two cases:
\begin{enumerate}
\item If we reflect  $z_{5}\raw -z_{5}$ (or equivalently $z_{6}\raw -z_{6}$), we get four fixed point in the fiber: fibering these points over the base $B$ one obtains the two divisors $D_5$ and $D_6$, that will be wrapped by O7-planes.
Given that $z_{5}\in \cO(3H)$,
the corresponding O7-plane/D7-brane setup provides the largest contribution to the D3-charge.
\item Let us now consider the involution $z_{4}\raw -z_{4}$: the fiber is inariant under it only when it degenerates to
\begin{align}
z_{5}^{2}&=z_6^2(a z_{4}^{2} + b   z_{6}^{4}) 
\end{align}
Unfortunately this singularity is inherited by the CY. Ignoring such a singularity, one may conclude that there is an O7-plane wrapping $D_{4}$, that does not touch the singularity because of the SR ideal.

\end{enumerate}

%

\bibliographystyle{JHEP}
\bibliography{Literatur}
\end{document}